\documentclass[11pt,a4paper]{article}
\pdfoutput=1
\usepackage[english]{babel}
\usepackage[latin1]{inputenc}
\usepackage[T1]{fontenc}
\usepackage{amsfonts,amsbsy,bm,euscript,mathrsfs}
\usepackage{amssymb,faktor,slashed}
\usepackage{color}
\usepackage[tbtags]{amsmath}
\usepackage[bookmarks=true,colorlinks=true,linkcolor=black,citecolor=black,urlcolor=black,bookmarksnumbered,linktocpage=true]{hyperref}
\usepackage{graphicx}
\usepackage{chngcntr}
\usepackage{makecell}
\usepackage{mathtools}

\usepackage{afterpage}

\usepackage{enumerate}

\usepackage[a4paper,text={170mm,257mm},centering]{geometry}

\numberwithin{equation}{section}

\makeatletter
\renewcommand\section{\@startsection {section}{1}{\z@}%
{-3.5ex \@plus -1ex \@minus -.2ex}%
{2.3ex \@plus.2ex}%
{\normalfont\large\bfseries}}
\renewcommand\subsection{\@startsection{subsection}{2}{\z@}%
{-3.25ex\@plus -1ex \@minus -.2ex}%
{1.5ex \@plus .2ex}%
{\normalfont\normalsize\bfseries}}
\makeatother

\expandafter\def\expandafter\bfseries\expandafter{\bfseries\ifmmode\else\boldmath\fi}
\expandafter\def\expandafter\mdseries\expandafter{\mdseries\ifmmode\else\unboldmath\fi}
\expandafter\def\expandafter\normalfont\expandafter{\normalfont\ifmmode\else\unboldmath\fi}

\providecommand{\href}[2]{#2}
\newcommand{\arxivlink}[1]{\href{http://arxiv.org/abs/#1}{[arXiv:#1]}}
\newcommand{\doilink}[2]{\href{http://doi.org/#2}{#1}}

\newcommand{\mathsym}[1]{{}}
\def\id{\protect{{1 \kern-.28em{\rm l}}}}
\def\be{\begin{eqnarray}}
\def\ee{\end{eqnarray}}

\def\ha{\tfrac{1}{2}}

\def\ci{\cite}

\def\ww{\Omega}

\def\z{\zeta}

\def\a{\alpha}
\def\b{\beta}

\def\g{\gamma}

\def\w{\omega}

\def\ta{{\tilde \a}}

\def\Tr{{\rm Tr}}

\def\l {\lambda}

\def\O{{\mathcal O}}

\def\foot{\footnote}
\newcommand{\rf}[1]{(\ref{#1})}

\def\no{\nonumber}

\def\la{\label}
\def\l{\lambda}

\def\p{\phi}
\def\r{\rho}

\def\varpi{{\rm w}}

\def\th{\theta}
\def\t{\tau}

\def\del{\partial}
\def\s{\sigma}
\def\eps{{\epsilon}}

\def\ed{\end{document}}

\def\iffa{\iffalse}

\def\ep{\epsilon}

\def\d{\delta}

\def\L{\mathcal{L} }

\def\sm{$\sigma$-model}
\def\sms{$\sigma$-models}
\def\ov{\over}

\def\Lie{\operatorname{Lie}}

 \def \h {{\rm h}}

\def\ddt{\frac{d}{d\t}}

\def\ed{\end{document}}

\def \cgg {\ccg}
\def\cg{c\,}

\def\e{\varepsilon}

\def\g{\gamma}

\def\ol{\underline}

\def\sms{$\s$-models\ }
\def\sm{$\s$-model\ }

\def\wt{\widetilde}
\def \wh  {\wt \rh}

\def \foot{\footnote}\def \ci{cite}\def \l {\lambda}\def \iffa {\iffalse}
 \def \ov {\over }\def \a  {\alpha} \def \ha {{1\ov 2}}
\def \ed {\end{document}}

\def \lab {\label}\def \la {\label}

\def \ov {\over}

\def \ha {{1 \over 2}}

\def \ci {\cite}

 \def \eps {\epsilon}

\def \rh {{\rm h}}
\def \GB {{$GB$}\ }
\def \cg {{\rm c}}

\def \ccg {{\cg_{_G}}} \def \ch {{\cg_{_H}}}
\def \cggg  {{\cg^2_{_G}}} 
 \def \ep {\epsilon}

\def \ww {{\rm w}}

  \def \bpsi {\bar \psi}
\def\cG{\ccg}

\def\ol{\overline}
\def \ha {\tfrac{1}{2}}

\def \B {{ I}}

\def \lh {h}
    \def \ta  {U}   \def \tb {V} 

\def \hij  {p_{ij} }  \def\Hij  {\chi_{ij}}

\begin{document}

\ \hfill{\small Imperial-TP-NL-2021-01 }

\vspace{0.5cm}

\vspace{2.5cm}

\begin{center}

{\Large\bf Integrability vs.  RG flow in  $G\times G$ and $G\times G/H$  sigma  models}

\vspace{1.5cm}
{
Nat Levine$^{a,}$\footnote{\ n.levine17@imperial.ac.uk} and
Arkady A. Tseytlin$^{a,}$\footnote{\ Also at the Institute of Theoretical and Mathematical Physics, MSU and Lebedev Institute, Moscow.
\\\hspace*{15pt} \ tseytlin@imperial.ac.uk}
}

\vspace{0.8cm}

{
\em \vspace{0.15cm}
$^{a}$Blackett Laboratory, Imperial College, London SW7 2AZ, U.K.
}

\end{center}

\vspace{1cm}

\begin{abstract}
We consider a class  of  2d  $\sigma$-models  on products of group spaces that   provide  new examples  of  a close connection   between  integrability and stability under the RG flow. We    first  study  the integrable $G \times G$   model   derived from the  affine Gaudin construction  (for which the 1-loop $\beta$-functions were found in \href{http://arxiv.org/abs/2010.07879}{arXiv:2010.07879})  and show that its condition of integrability  is   preserved also  by the 2-loop  RG flow.  We  then   investigate the RG flow   in the gauged $G\times G/H$  model, in particular  the integrable  $T^{1,1}$    model  found in \href{http://arxiv.org/abs/2010.05573}{arXiv:2010.05573}. We also  construct   a   new   class of integrable $G\times G/H$ models in the case  when the subgroup  $H$ is abelian. In the  simplest case of $G=SU_2,\ H=U_1$  this  leads  to an integrable $\sigma$-model on the   $T^{1,q}$  space  (with a particular $B$-field). This  model is  also shown to be stable under the 2-loop  RG flow, and we  relate this   property to its  invariance  under T-duality in an isometric $U_1$ direction. This  $T^{1,q}$ model   may be interpreted as an integrable  deformation of the GMM model  (of two coupled WZW  theories  with generic levels) away from the conformal point.

\end{abstract}
  
\newpage 
\tableofcontents

\setcounter{footnote}{0}
\setcounter{section}{0}
\begin{center}
{\large \bf
 }
\end{center}

\section{Introduction}

It is  
 expected  that classically  integrable 2d \sms   should be   stable  under the  renormalization group  flow, 
 the intuition  being that hidden symmetries 
will constrain
 the  RG evolution. 
Constraints on coupling constants   required for integrability should thus be RG-invariant. 
At the leading 1-loop order, this has been observed for some time (see, e.g.,  \cite{intRG}). It was recently found 
 on various examples \cite{HLT1,HLT2}  that the RG stability 
for integrable theories 
  extends also to 
  higher-loop orders (provided the classical 
 actions  are supplemented by particular finite counterterms or if RG evolution is considered on a larger  configuration space). 

The aim of this paper is to explore the  connection between integrability and the RG flow  on 
some new examples ---   integrable $G\times G$ and $G\times G/H$ models that were derived from  the affine Gaudin construction \cite{DLMV, ABL}.  These models may be viewed as    generalizations  of the PCM$_k$,\foot{In our 
conventions  
    the action   is $S= \frac{1}{4\pi\a'} \int d^2 \xi \, \L$  
with
  the "string"  notation  for the loop counting parameter $\a'\equiv \hbar$  that  may be set to 1  in some of the equations
below. We also  use $\del_\pm = \del_0 \pm \del_1$.}
\be
\L_{_{{\rm PCM}_k}} = \L_{_{{\rm PCM}}}  + k \, \L_{_{\rm WZ}}(g)\ , \qquad  
\L_{_{{\rm PCM}}}  = -\ha \, {\rh} \, \Tr[ J_+ J_- ]\ ,   \qquad J \equiv g^{-1} dg \ , \ \ g\in G \ , \la{Pk}
\ee
i.e.\   the principal chiral model    (with inverse coupling $\rh$)   with  the WZ term  (with "level" $k$).
The conformal WZW model is obtained at the special points $\rh=\pm k$.

The PCM$_k$   admits  various  integrable deformations (see, e.g., \cite{klim, balog,DMV}), which have been  interpreted 
  \cite{DLMV} as  particular  cases  of   integrable affine Gaudin models. 
The  affine Gaudin construction also produces    natural generalizations  of the  PCM$_k$  to integrable models on products of group  spaces $G^N = G\times \ldots \times G$ \cite{DLMV}. 

Here  we shall  consider  a  subclass of  such  models  defined  by 
\begin{align}
&\L = -\ha \, \rho_{ij} \, \Tr[ J^{(i)}_+ J^{(j)}_- ] + k_i  \, \L_{_{\rm WZ}}(g^{(i)}) \ ,  \qquad \qquad \la{cou}\\
&J^{(i)}\equiv {g^{(i)}}^{-1} d g^{(i)} \ , \qquad \qquad  g^{(i)} \in G^N  \ , \qquad i=1,\ldots,N \ . \no
\end{align}
We denote by  $J^{(i)}$   the Maurer-Cartan 1-form  corresponding to $i$-th copy of  $G$  and 
 $\rho_{ij}$ is a {\it constant }  coupling   matrix (summation over repeated  $i,j$ is assumed). 

The PCM$_k$ \rf{Pk} corresponds to the special case $N=1$ (with $\rho_{11}=\rh$ and $k_1=k$), and is integrable for any values of its couplings. However, for $N>1$,   the model \rf{cou} is  classically integrable only for \textit{special}  couplings $(\rho_{ij}, k_i)$
 that correspond to  
 the  affine Gaudin models \cite{DLMV}. 
These are
selected  as the solutions 
 of certain polynomial equations.
We will focus  on the first  non-trivial case of $N=2$, i.e.\  on  $G\times G$  models.

 As we shall find in  Section 2,
  the 
classical integrability condition for 
 $G\times G$ theories \rf{cou} 
is 
 automatically stable 
 under the  \textit{2-loop} RG flow 
 in
 a particular  subtraction scheme (extending the 1-loop results of \cite{today}). 
 Here the  2-loop stability is  obtained without the need for any  finite counterterms.

\

The model  \rf{cou}  
is
 a special case   of the 2d 
 $\s$-model
\be \lab{13}
S = \frac{1}{4\pi\a'} \int d^2 \xi \, \L = \frac{1}{4\pi\a'} \int d^2 \xi \, \big[ G_{mn}(x) + B_{mn} (x)\big] \,  \del_+ x ^m \del_- x^n \ .
\ee
This is a "two-coupling" theory, so the  2-loop $\beta$-functions for $(G,B)$  
generally 
depend on a  
choice of 
a renormalization scheme \ci{mt}. There exists a  special 2-loop   scheme  \cite{Bos,mt}
that effectively  treats 
$G_{mn}$ and $B_{mn}$  as symmetrically as possible (with the respective 
$\beta$-functions   being the symmetric and antisymmetric parts of a single tensor expression). 
  We shall refer to this $G$-$B$ symmetric scheme   as  the
  "\GB scheme". 
Explicitly, in this  scheme one finds for the 2-loop $\beta$-functions 
 \ci{mt}  (see also \cite{Hull})\foot{Here $\tau$ is the RG parameter.
 In general, the $\beta$-functions   may   contain also   diffeomorphism and  $B$-gauge transformation 
terms  corresponding to  freedom of field renormalizations and   shifts of  the Lagrangian by 
total derivatives 
 depending on RG scale. 
We omit these terms since they  
automatically 
   vanish in the examples  considered below 
 due  to  manifest global $G_L \times G_L$  symmetry.}
\be \la{BMT} \begin{aligned}
\ddt(G_{mn} + B_{mn}) &= \a'\,  \b^{(1)}_{mn} + \a'^2 \,  \b^{(2)}_{mn} + \ldots \\
&= \a' \, \widehat R_{mn} +  \a'^2 \, \tfrac{1}{2} \Big[\widehat R^{klp}{}_n \widehat R_{mklp} - \tfrac12 \widehat R^{ l p k}{}_n \widehat R_{mklp} 
+\tfrac12 \widehat R_{k mn l} H^{k pq} H^ {l}_{\ pq} \Big] + \ldots \ . 
\end{aligned} \ee
Here $H_{mnk}=3 \del_{[m} B_{nk]}$   and  $\widehat R$ is  the  curvature  of the generalized connection $\widehat\Gamma^k{}_{mn} = \Gamma^k{}_{mn}(G) - \tfrac12 H^k{}_{mn}$.
Applied  to the   case of the PCM$_k$  in  \rf{Pk},  the expression in  \rf{BMT}  gives (here we set $\a'=1$)
\be \ddt{\rm h} = \ccg\, (1-\tfrac{k^2}{\rh^2})\, \Big[ 1 +  \tfrac{\ccg}{2\rh} (1-  \tfrac{3\, k^2}{\rh^2}) \Big] \ ,  \qquad \qquad   \ddt{k}=0  
 \ , \la{Pkb}\ee
so that  the  position  of the WZW  fixed point $\h= \pm k$ remains unchanged at the 2-loop order. 
The  2-loop PCM$_k$ $\beta$-function \rf{Pkb} was 
found in  \cite{Bos}  
using a scheme equivalent (at the 2-loop level)
 to the one of \cite{mt} that leads to \rf{BMT}.\foot{To recall, part of the scheme freedom
comes from the  prescription of how one treats  the  antisymmetric  2d tensor  $\e^{ab}$  appearing  in the $B$-term in \rf{13} 
 in dimensional regularization. Ref.  \cite{Bos}  used  't Hooft-Veltman  prescription of treating $\e^{ab}$ as effectively 2-dimensional. 
 In  \cite{mt} it was  assumed  that, in $d=2+\ep$ dimensions,  $\e^{ab}\e_{cd} = 
  f(\ep) ( \delta^a_c \delta^b_d - \delta^a_d \delta^b_c ) $  where   $f= 1 + f_1 \ep +...$,    and then the  \GB scheme  corresponds to 
the choice  $f_1= -1$.  As noted in \cite{Bap}, the scheme used in \cite{Bos}   
is equivalent (at least at the 2-loop level)  to  
$f(d) = \frac{1}{ d-1}= 1 - \ep + ...$, i.e.\  to the choice $f_1= -1$
 \ci{mt}
of the \GB scheme \rf{BMT}.}

The  \GB   scheme    is naturally  "adapted" to the  vicinity of the WZW  conformal point: 
the derivative $\del_\rh \b_\rh\big|_{\rh=k}$ of the $\beta$-function for $\rh$ at the fixed point
correctly  reproduces \cite{Bos}  the  anomalous  dimension of  the $\Tr (J_+J_-)$ operator (PCM  Lagrangian)
as computed \cite{KZ} using the  underlying  infinite dimensional 
Kac-Moody  symmetry  of the WZW   model.   Thus   this scheme is 
apparently 
 consistent  with the preservation of the KM symmetry  in 
  the vicinity of  
the conformal point. 

  It is then natural to expect that this scheme   should also  play  a special role   in  a more general  class  of 
 integrable  models \rf{cou}   containing   WZW  models as  special    limits,\foot{Similar logic was recently used 
    in \cite{shifman} in the  discussion of the 2-loop RG   evolution  of a   "squashed"  $SU_2$  variant  of  
   PCM$_k$.} 
   and should  facilitate  preservation   of the  hidden integrable structure of these models at the quantum level.  
We will indeed  see    evidence for this below:  the 
classical integrability  conditions for the $G\times G$ model  \rf{cou} 
will be automatically  preserved by the 2-loop   RG evolution 
 provided  one uses the $\beta$-functions in the \GB scheme  \rf{BMT}. 

\

We shall also  study, 
in Section 3, 
 a  gauged  analog  of the  models \rf{cou}  defined  on a coset space $G\times G/H$.
 This theory,  
  which
 was  recently derived from  affine Gaudin models  in  \cite{ABL},
may be   viewed as a
    generalization  of the standard  $G/H$ symmetric space $\s$-model,   
    also including 
 WZ terms.
    For these $G\times G/H$ theories to be gauge invariant, the  corresponding 
    couplings must satisfy certain linear relations. In addition, for a gauge invariant   model to be  classically integrable,
    the couplings  should  further satisfy certain polynomial relations \cite{ABL}. 
        
    We will 
    compute  the RG flow for these 
     integrable $G\times G/H$ theories, finding that  they  are stable under the 
      1-loop RG flow.  
      However, at the 2-loop level, 
RG stability 
does  not
 automatically   
 arise  and,  in general,   
          requires   certain 
      finite redefinitions of the couplings. These  are  
       equivalent to adding specific  finite counterterms,         
which  may  be  interpreted  as  required for preservation of integrability  
at the quantum level
(this is analogous to what was observed  on other   examples in     \cite{HLT1,HLT2}). 
       There are still a few special  cases, in particular the integrable $T^{1,1}$ model of  \cite{ABL},   that are  automatically stable 
       under the 2-loop RG flow (see Section 3.2). 

\

\sloppy
In Section 4 we shall present a new  integrable \sm with 
target space  metric  
${T^{1,q} = SU_2 \times SU_2/U_1}$ \ci{romans} 
   and a particular $B$-field. The model admits as a special limit 
 the  conformal GMM model  with unequal levels \ci{GMM,PZT}.
Our central observation is that, in the case of the  subgroup $H$ in $G\times G/H$  being 
 abelian, the gauge invariance conditions of  \cite{ABL}  are too restrictive and  there is also a second 
 "branch" 
  of gauge invariant theories. This allows a natural generalization of the integrable $T^{1,1}$ model of \cite{ABL}  to $T^{1,q}$ 
  with a general  parameter 
$q$. We demonstrate that the resulting $T^{1,q}$ model is classically integrable, admitting a Lax representation. We  observe 
 that the $T^{1,q}$ model is self-dual  under T-duality in one  isometry direction, and argue that this  property 
forces it to be stable under the RG flow. 
We verify this fact  explicitly  by computing  the corresponding  2-loop RG flow of the two  coupling constants.

 \

A few concluding remarks  will be made in  Section 5. 
In Appendix \ref{A}  we shall  discuss  the integrability conditions for the $G^N$ model \rf{cou}. 
 In Appendix \ref{B} we shall provide the explicit formulae for the 2-loop $\beta$-functions of the $G\times G$ and $G\times G/H$ models
  and explain how they were derived.

\section{$G\times G$ models \la{2}}

As was mentioned  in the Introduction, the  $G^N$ model \rf{cou} is  classically integrable for \textit{special} values of its couplings $(\rho_{ij}, k_i)$ satisfying certain polynomial relations, which originate from the  affine Gaudin  construction
 \cite{DLMV}. For such values of the couplings the model  admits a  Lax connection of the form
\be
L_+ = \a_i \,  J^{(i)}_+ \ , \qquad\qquad  L_- = \b_i  \, J^{(i)}_- \ , \la{Lan}
\ee
whose flatness condition, $F_{+-}(L) \equiv \del_+ L_- - \del_- L_+ + [L_+, L_-] =0$, is equivalent to the equations of motion
following from \rf{cou}.  Moreover, the affine Gaudin construction guarantees that the Poisson brackets of the Lax matrix $L_\s = \ha (L_+ - L_-)$ can be written  in a   `twist'  form,  i.e.\   a special form of the standard non-ultralocal $r/s$ Poisson bracket  \cite{maillet}. This 
implies 
the existence of  a tower of conserved  commuting higher-spin charges 
\cite{LMV}. 

Below  we shall consider  the simplest $N=2$ case of  the $G^N$ model  \rf{cou} for a simple Lie group $G$.  
We shall parametrize the $2\times 2$ matrix $\rho_{ij}$ in \rf{cou} in terms of the 4 components $s,t,u,b$ as follows 
\begin{align} 
\L = -\ha \,\begin{pmatrix}
s & t+b \\ t-b& u
\end{pmatrix}_{ij} \, \Tr[ J^{(i)}_+ J^{(j)}_- ] + k_i  \, \L_{_{\rm WZ}}(g^{(i)}) \ . \la{par}
\end{align}
Then the  affine Gaudin condition for   integrability is the vanishing of  a  cubic polynomial \cite{DLMV,today},
\be
f(s,t,u,b ,k_1,k_2) \equiv -t \, (s+t)\, (t+u)+b^2\, (s+t+u)+t \, k_1\,  k_2 + b \, (u\,  k_1 - s\,  k_2) = 0  \la{cond} \ .
\ee

Let us  note that the 
affine Gaudin conditions for integrability 
(e.g.\ \rf{cond} in the $N=2$ case) are certainly sufficient for integrability. However, it is not  a priori  clear  if they are  necessary,  
  since there could also be integrable theories of the form \rf{cou} that are unrelated to the affine Gaudin  construction 
of   \cite{DLMV}.
In Appendix A we   presented a  check that the condition \rf{cond} is also \textit{necessary} for the integrability
of the $G\times G$ model \rf{par},  assuming the natural ansatz \rf{Lan} for the corresponding  Lax connection.

\subsection{RG flow in   $G\times G$ models}

The general $G^N$ model \rf{cou} has  global $(G_L)^N \times G_R$ symmetry acting as
\be
g^{(i)} \to u_L^{(i)} g^{(i)} u_R\ , \qquad\qquad   (u_L^{(i)}, u_R) \in (G_L)^N \times G_R \ . \la{GGs}
\ee
In fact,  \rf{cou} is  the most general 2-derivative  local Lagrangian  having  this symmetry. 
 This implies  that only $\rho_{ij}$ can  run  under the RG flow
(with the WZ parameters $k_i$  not renormalized as usual).\foot{As in the PCM$_k$ case,  the  RG invariance of $k_i$ 
  follows from the fact that 
the corresponding field strength $H=dB$ is covariantly constant.}

Starting with the \sm couplings $(G_{mn},B_{mn})$   corresponding to the   $G \times G$ model 
\rf{par}   and  computing the 
corresponding 2-loop $\beta$-functions   in the \GB  scheme  \rf{BMT}, 
we find  
\be\la{2c} \begin{aligned}
 & \qquad\qquad \ddt \rho_{ij}  = \a' \, \b_{ij}^{(1)} + \a'^2\,  \b_{ij}^{(2)} + \ldots \ , \\
 \b_{ij}^{(1)} = {}&{} \cgg \, F^{(4)}_{ij}(s,t,u,b ,k_1,k_2) \ ,  \qquad  
  \b_{ij}^{(2)} = {\rm c}_{_G}^2 \, (su-t^2)^{-5} \, F^{(9)}_{ij}(s,t,u,b ,k_1,k_2) \ , 
\end{aligned} \ee
where the matrices $F^{(4)}, F^{(9)}$ are  homogeneous polynomials of degrees 4 and 9 in their arguments 
and $\ccg$  is the dual Coxeter number of  the group $G$, as in  the PCM$_k$ case in \rf{Pkb}.
The   explicit  expressions 
for $F^{(4)}, F^{(9)}$
 are given in  Appendix \ref{GGapp}    and also in some special cases   below.

Remarkably, despite the complicated   expressions  for the   $\beta$-functions,  one is able to verify   that 
  the integrability condition \rf{cond} is, in fact, preserved by the 2-loop RG flow:
\be { d f\ov d \tau} \Big|_{f=0} = \big( \a' \, \b_{ij}^{(1)} + \a'^2 \, \b_{ij}^{(2)} + ... \big) \, { \del f \ov \del {\rho_{ij}}  }\Big|_{f=0} = \a'\times 0+ \a'^2\times 0 + ... \ .  \la{GGstab}
\ee
The vanishing of the 1-loop $\O(\a')$ term in \rf{GGstab} was already established in \cite{today}, and the vanishing of the 2-loop 
 term is a  new  non-trivial  result. Let us stress   that 
 this property of 
 the integrability condition \rf{cond}
not being deformed  at the 2-loop level is specific to the 
  \GB   scheme \rf{BMT}. 

\subsection{Some special cases  \la{sc}}
Let us  consider some particular examples of the  integrable $G\times G$ models \rf{par},\rf{cond}.

\subsubsection{$\rho_{21}=0$ and the $G\times G$ model related to $\l$-model}
The most general integrable model with $\rho_{21}=0$  corresponds to the  following choice  of the parameters in \rf{par}
(this case was also considered in Appendix C of \cite{today})
\be
b=t \ , \qquad u = -k_2 \ .
\ee 
After the redefinition $  (g^{(1)},g^{(2)})\equiv(g,\wt{g}^{-1})$, the corresponding Lagrangian \rf{par}  
depending on $s,t, k_1, k_2$ 
may be written as 
  (cf. \rf{Pk}) 
\begin{align}
\L ={} &\big[ s \L_{_{\rm PCM}}(g) +k_1  \L_{_{\rm WZ}}(g)\big]  -  k_2 \big[  \L_{_{\rm PCM}}(\wt g) + \L_{_{\rm WZ}}(\wt g)
\big]+ t \, \Tr
\big[ J_+(g) K_-(\wt g)\big]  \ , \la{1}
\end{align}
where 
$J= g^{-1} d g,   \  K(\wt g) = d \wt{g}\,  \wt{g}^{-1}$. 
This is   just a  PCM$_{k_1}$    and  WZW model (with level $-k_2$) 
 coupled  via   the $J_+(g) K_-(\wt g)$ term. 
In this case the global $G_R$ symmetry in \rf{GGs} is enlarged to a chiral symmetry  $G_R(\xi^+)$,\foot{We denote by $G(\xi^+)$  right $G$  multiplications 
 depending on  light-cone coordinate $\xi^+= \ha (\xi^0 + \xi^1)$.\la{chir}}
\be\la{29}
(g,\, \wt{g}) \to (u\, g\, v, \ v^{-1} \wt{g}\,  w(\xi^+)) \ , \qquad \qquad (u,v,w(\xi^+))\in G_L \times G_L \times G_R(\xi^+)   \ .
\ee
These   symmetries protect the structure of \rf{1}  under
 renormalization so  that only the parameters $s$ and $t$ are expected to run  with the RG scale.  Indeed,  in this case the RG equations  \rf{2c}  take the following explicit  form\foot{As in \rf{Pkb},   here we set 
the loop counting parameter $\a'$ to  be 1. 
The 1-loop terms in \rf{sr},\rf{tr} match those in \cite{today} (after  reversing the sign  of the WZ terms $k_i \to - k_i$  to match the conventions).}
\begin{align}
\ddt s {}&{}= \frac{\cgg\, (s-k_1)  }{\left(k_2 s+t^2\right)^2} \Big[k_2^2 (k_1+s)+4 k_2 t^2-2 t^3\Big] \no \\
&\quad + \frac{\cggg\, (s-k_1) }{2 \left(k_2 s+t^2\right)^5} \Big[ 2 k_2 t^5 \left(38 t^2-11 k_1t +2 s^2+41 s t\right) \la{sr} \\
&\quad 
 +2 k_2^3 t^3 \left(-8 k_1 s-42 k_1 t+9 k_1^2-5 s^2+18 s t\right) 
  -2 k_2^2 t^4 \left(-7 k_1 s-46 k_1 t+s^2+48 s t+28 t^2\right)\no\\
&\quad 
+k_2^5 (k_1+s) \left(s^2-3 k_1^2\right)+2 k_2^4 t^2 (3 k_1+2 s) (3 s-5 k_1)-4 t^7 (5 s+6 t)\Big]\no  \ , \\
\ddt t{}&{} =\frac{ \cgg\, t (t-k_2)  }{\left(k_2 s+t^2\right)^2} \Big[ k_2 (k_1-s)+2 t (s+t)\Big] \no \\
&\quad + \frac{\cggg\, t (t-k_2)}{2 \left(k_2 s+t^2\right)^5}  \Big[4 t^5 \left(t (4 t-k_1)+5 s^2+10 s t\right)-k_2^4 (s-k_1) \left(s^2-3 k_1^2\right)\la{tr} \\
&\quad 
+2 k_2^2 t^2 \left(s^2 (28 t-3 k_1)+2 s t (13 t-16 k_1)+6 k_1 t (k_1-4 t)+3 s^3\right)\no\\
&\quad 
-2 k_2 t^3 \left(-k_1 t (13 s+19 t)+2 s^3+31 s^2 t+45 s t^2+10 t^3\right) 
-2 k_2^3 t (s-k_1) (5 s t-3 k_1 (s+4 t))\Big] \ . \no 
\end{align}
At the obvious  fixed point   $s=k_1, \ t= k_2$, the model  \rf{1}   becomes  \cite{today}  
the sum of two decoupled WZW models,
$\L = (k_1+k_2) \L_{_{\rm WZW}} (g) - k_2 \L_{_{\rm WZW}}(g \wt g)$.
 As discussed in \cite{today}, the fixed points are all decoupled WZW models of this type.
  The RG trajectories either interpolate between such WZW-type fixed points or 
flow to them in the IR from the  asymptotically free  UV fixed  point $s, t \to \infty$.

An  interesting  special case of \rf{1}  is $s=k_1=- k_2\equiv - k' $, when it becomes 
\be
\L =- k'  \Big(  \L_{_{\rm WZW}}(g)  + \L_{_{\rm WZW}}(\wt {g})  - \l' \, \Tr\big[ J_+(g)  K_-(\wt g)\big]  \Big)  \ , \qquad \quad  \l' \equiv k'^{-1}t \ .\la{212}
\ee 
This particular $G\times G$ model  appears from the "tripled"  version \cite{HLT2} 
of the $\l$-model \ci{Sfetsos:2013wia}
after removing the  decoupled WZW part.  It is also 
 a special case of the "doubly $\l$-deformed" model of \cite{sfetsos}. 
 Here  the  $\beta$-functions   \rf{sr},\rf{tr} reduce to just $\l'$ running as 
\be\la{213}
\ddt \l' = \frac{2\cgg\, \l'^2}{k' (1+\l')^2} + \frac{4{\rm c}^2_{_{G}}\,  \l'^4(1-2\l')}{k'^2 (1-\l')(1+\l')^5} \ .
\ee
This is the 2-loop $\beta$-function \cite{HLT2}
for the $\l$-model 
based on  
the group $G$ 
with  parameters $(k, \lambda)$  related to $(k',\l')$ as 
$k' =k+2 \cgg$, $\l' = \tfrac{k}{k+2\cgg}\l^{-1}$.\foot{See also \cite{s19}. For the 1-loop beta functions of the $\l$-models based on $G$ and $G/H$, see \cite{ss} and \cite{ah} respectively.}

\subsubsection{$k_1=k_2=0$}
 Setting the WZ levels to zero, $k_1=k_2=0$,  
 the integrability condition \rf{cond}  implies that 
\be \la{214}
b = b(s,t,u) \equiv \Big[{ t(t+s)(t+u)\ov t + s+u}\Big]^{1/2} \ . \ee
 We thus obtain 
 from  \rf{par} 
 an integrable  $G \times G$ model  with  3  independent couplings $s,t,u$,
\be
\L = -\tfrac{1}{2} \begin{pmatrix}
s & t+ b(s,t,u) \\
t- b(s,t,u) & u
\end{pmatrix}_{ij} \Tr[ J^{(i)}_+ J^{(j)}_- ]   \ .
\ \la{matb}
\ee
Since $k_i$  do not run, this special case of the model \rf{par},\rf{cond}  should also  be stable under the RG flow, i.e.\ 
\rf{matb}   should be renormalizable  with only $s,t,u$  running. 
 Indeed, 
using for convenience the redefined 
couplings  $(s,t,u)\to (x,y,z)$ with
 \be\la{216}
x = s+t+u \ , \qquad y = \frac{s}{t} \ , \qquad z = \frac{u}{t} \ ,\la{rc}
\ee
 the 2-loop $\beta$-functions \rf{2c}  become 
 \begin{align}
&\qquad \ddt x = 2 \cgg - \frac{ \cggg }{2 x (y z-1)^3} \  \Big[ 16 +32 (y + z) + 16 ( y^2 + z^2) + 88 y z + 68 y z (y+z)  \no \\
 & \qquad \qquad \qquad  \qquad \qquad \qquad\qquad + 12 y z (y^2+z^2 + 5 y z )   + 8 y^2 z^2 (y+z)    - y^2 z^2 (y +  z)^2  \Big]\ ,    \la{br1}\\
&\qquad \ddt y =F(x;y,z)\ , \qquad \qquad \ddt z = F(x;z,y) \ , \la{218} \\
& \qquad  F(x;y,z)\equiv  \frac{  \, y  (y+1) (y+2) }{ \,  (zy-1)^2} \ \bigg(  \frac{ \cgg}{x} \big[ 1-zy- 3(z+1)^2 \big] \  \no \\
&\qquad \qquad \ \   
 -\frac{ \cggg}{2x^2} \frac{1}{ (z y-1)^3 }\Big[  -z^6 y^2- y^6 (3z  y-38z-44) 
 +2 z (y+1) (26 zy+101y+58) +20 (y+1)^2\no \\
& \qquad\qquad  
\ \ \  -z^4 (y (3 y ((y-14) y-100)-296)-38)  
-z^3 (y (y (y ((y-4) y-178)-728)-708)-152) \no\\
& \qquad\qquad\qquad \qquad \ \ \  +2 z^2 (y+1) (y (y (5 y+89)+262)+105)  \Big] \bigg)\ . \la{br3}
\end{align}
The obvious  symmetry  between $s$ and $u$ in \rf{matb}  is  translated into  the symmetry of the RG equations under 
$y\leftrightarrow z$.

The fact  that 
these 2-loop $\beta$-functions are much simpler than  
 the general (not necessarily integrable) case  of \rf{2c} 
 (see also  
 Appendix  \ref{GGapp}) 
 suggests that a   substantial simplification  happens 
 upon specifying the 
 couplings to be  at  the integrable locus $f=0$
 in 
  \rf{cond}
  (this   was already observed at the 1-loop order in \cite{today}).

\section{$G\times G/H$ models \la{gu}}

Let $H$ be a subgroup  of $G$ such  that $G/H$ is a symmetric space (we assume that both $G$ and $H$ are simple real Lie groups).  Then the 
 Lagrangian for the gauged $G\times G/H$ model  of \cite{ABL} takes the form\foot{Our conventions in 
\rf{couH} are related to the ones of  \cite{ABL}  by  $r_{ij} \to 2 \rho^{(0)}_{ij}$, $\rho_{ij} \to 2 \rho_{ij}^{(1)}$ and
 the  opposite sign  for 
the WZ terms, i.e. $k_i \to - k_i 
$.}
\begin{align}
&\L = -\ha \, \rho_{ij} \, \Tr[ P^{(i)}_+ P^{(j)}_- ] -\ha \, r_{ij} \, \Tr[ \B^{(i)}_+ \B^{(j)}_- ] + k_i  \, \L_{_{\rm WZ}}(g^{(i)}) \ , \la{couH} \\
& P^{(i)} = P_{{G/H}}  J^{(i)} \ , \qquad   \B^{(i)} = P_H J^{(i)}\ ,  \ \ \qquad 
 J^{(i)} =  {g^{(i)}}^{-1} d g^{(i)}  \ . 
  \la{31}
\end{align}
Here  $P_{{G/H}}  $ and $P_H$ are projectors to the corresponding  parts of
 the algebra of $G$,   and 
$\r_{ij}$, $r_{ij}$ are  
 constant $2\times 2$ matrices.  
The global symmetry 
consists of 
 left multiplication $G_L\times G_L$, as well as the discrete  $\mathbb{Z}_2$  corresponding to
 the symmetric space structure of $G/H$.
The  action  for \rf{couH}
 is required  to be gauge invariant under the local  right action  by an element of $H$ (acting 
 the 
 same 
 on  both $g^{(i)}$) 
\be
g^{(i)} \to g^{(i)} \ww\ , \qquad \qquad  \ww(\xi^+, \xi^-)\in H \ .  \la{gI}
\ee
 For general choices of $G$ and $H$,  
   gauge invariance  imposes the 
   linear  
   constraints  \cite{ABL}\foot{The   special case of  abelian $H$   will be   
 discussed below in Section 4.}
\be
 k_1 = - k_2\equiv k  \ , \qquad \qquad r_{ij} = \begin{pmatrix} r  & - r - k \\ -r +k & r  \end{pmatrix} \ . \la{bb1}
\ee
The remaining free parameters  of the gauge invariant model 
are then $r$, $k$ and  the $2\times 2$ matrix $\rho_{ij}$.

Requiring integrability imposes  further constraints  which,
as for the
$G^N$ models \rf{cou}, 
can be  obtained from the affine Gaudin  construction. 
The following parametrization of  the 6 constants $r,k, \rho_{ij}$ 
 in terms of   4 parameters $K,x,\z_+,\z_-$ was shown  in \cite{ABL} to be   sufficient for   integrability
\begin{align}
&r=r_{11} = r_{22} = K \frac{\z_-^2-\z_+^2}{(1-x^2)^2} \ , \quad r_{12} = 2K \frac{(1-\z_+^2)(x^2-\z_-^2)}{(1-x^2)^3}\ , \quad r_{21} = -2K \frac{(1-\z_-^2)(x^2-\z_+^2)}{(1-x^2)^3} \ , \la{iH1} \\
&\r_{11} = K\frac{1-2\z_+^2 +\z_-^2 \z_+^2}{(1-x^2)^2} \ , \ \ \qquad \qquad  \r_{12} = x \, r_{12}= 2 K \frac{x (1-\z_+^2)(x^2-\z_-^2)}{(1-x^2)^3} \ , \\
& \r_{21} = x^{-1} \, r_{21} = -2K \frac{(1-\z_-^2)(x^2-\z_+^2)}{x (1-x^2)^3}  \ , \qquad \qquad \r_{22} = K\frac{x^4-2\z_+^2 x^2 +\z_-^2 \z_+^2}{x^2(1-x^2)^2} \ , \\
&k=k_1 = -k_2 = - K \frac{2x^2 +2\z_-^2\z_+^2-(1+x^2)(\z_-^2+\z_+^2)}{(1-x^2)^3} \ . \la{iH4}
\end{align}
This parametrization  is simply equivalent to the gauge invariance conditions \rf{bb1} 
combined with the   two extra polynomial integrability conditions
\be \begin{aligned} 
f_1 \equiv {} &{} r^2 - k^2 - \r_{12} \r_{21}=0  \ , \\
 f_2 \equiv {}&{}  (r - k)^4  \r_{12} + (r - k)^2 (r -k -  2 \r_{11}) (r -k - 2 \r_{22})  \r_{21}    \\
&{} \qquad \qquad \ \ \ 
 - 2(r-k)   (\r_{11} + \r_{22})  \r_{12} \r_{21}^2 +  (\r_{12} +  \r_{21})\r_{12}  \r_{21}^3  =0   \ . 
\end{aligned}  \la{condH} \ee
Two  simple solutions of  these   conditions are found by setting 
 $r=k$ (i.e. $r_{21}=0$  in \rf{bb1})     and   
either 
    $\r_{21}=0$  or   $\r_{12}=0$.

\subsection{RG flow in  $G\times G/H$ models}

The structure of the gauge invariant $G\times G/H$   action  \rf{couH},\rf{bb1} is protected 
by the  right $H$ gauge symmetry \rf{gI}  and the global $G_L\times G_L $ and  $\mathbb{Z}_2$ symmetry.
This rules out 
all
counterterms    
except those
  corresponding  to   renormalizations 
  of the 6 couplings $r,k,\rho_{ij}$ (of which $k$ is not renormalized as usual). 
Let us  parametrize $\rho_{ij}$ as in \rf{par},
\be\la{rrr}
\rho_{ij} = \begin{pmatrix}
s & t+b \\
t-b & u 
\end{pmatrix} \ .
\ee 
Computing the  $\beta$-functions \rf{BMT} corresponding to the  \sm couplings $(G_{mn},B_{mn})$   for the model \rf{couH},\rf{bb1},\rf{rrr}, 
we find for  the 1-loop $\beta$-functions of the 5 running  couplings
\begin{align}
& \ddt \lh_p  = \a' \b_{\lh_p}^{(1)} \ , \qquad \qquad \lh_p\equiv (r,s,t,b,u) \ , \\
&\b_r^{(1)} = \frac{\ccg- \ch}{(t^2 - s u)^2} \Big( r^2 s^2 - 2 b^2 t^2 - 2 r^2 t^2 + 2 t^4 -  2 b^2 s u - 2 s t^2 u + r^2 u^2 \no \\
   &\qquad \qquad \qquad \qquad + 4 b s t k + 4 b t u k - s^2 k^2 - 2 t^2 k^2 -  u^2 k^2 \Big) 
    + \ch \big(1-\tfrac{k^2}{r^2}\big) \ , \la{1H}\\
&\b_s^{(1)} = \frac{\ccg}{r (t^2 - s u)} \big[ b^2 s - s t^2 + r^2 u + 2 r (t^2 - s u) -  2 b t k + u k^2 \big]  \ , \\
  &\b_t^{(1)}= \frac{\ccg}{r (t^2 - s u)} \big[-b^2 t + b (s + u) k + t (r^2 - s u - k^2) \big] \ , \\
  & \b_b^{(1)} = \frac{\ccg}{r (t^2 - s u)} \big[-b (t^2 + s u) + t (s + u) k \big] \ , \\
  &\b_u^{(1)} =  \frac{\ccg}{r (t^2 - s u)} \big[ r^2 s + b^2 u - t^2 u + 2 r (t^2 - s u) - 2 b t k + s k^2 \big] \ . \la{fH}
\end{align}
 We observe   that the integrability conditions \rf{condH} are 
 stable under the 1-loop RG flow \rf{1H}--\rf{fH},
\be\la{317}
{\del  f_a\ov \del \tau} \Big|_{f_1=f_2=0} = \a'  \beta_{\lh_p} ^{(1)} {\del f_a\ov\del \lh_p} \Big|_{f_1=f_2=0} + \O(\a'^2)  = 0 + \O(\a'^2) \ , \qquad\qquad  a=1,2 \ .
\ee
However,  it turns out   that (as for   some  examples discussed in \cite{HLT1, HLT2})
 this property of RG stability does not, in general,  extend to the 2-loop order. 
Computing the 2-loop $\beta$-functions  for the model \rf{couH},\rf{bb1} in the \GB scheme \rf{BMT} (given explicitly in Appendix \ \ref{GGHapp}), 
we find that  
 the subleading correction to \rf{317} is non-zero at general values of the couplings, 
\be\la{318}
 \beta_{\lh_p} ^{(2)} {\del f_a\ov\del \lh_p} \Big|_{f_1=f_2=0} \not= 0\ ,  
 \qquad\qquad  a=1,2 \ .
\ee 
Moreover, we checked that 
 \rf{318} is also non-vanishing  in  
arbitrary  covariant  2-loop subtraction schemes.\foot{More precisely, we considered arbitrary subtraction schemes related to the \GB scheme \rf{BMT} by covariant redefinitions of $G_{mn}$ and $B_{mn}$.}

As in other examples \cite{HLT1,HLT2}, one may expect to restore the property of  RG stability at the 2-loop order by adding certain finite quantum $\a'$-corrections to the target space geometry.
Because of the global and local symmetries, the only possible corrections would 
correspond to   redefinitions  $\lh_p \to \bar \lh_p$  of the couplings  $\lh_p= (r,s,t,b,u)$,
\be 
 \lh_p=\bar  \lh_p + \a' Q_p(\bar \lh) + \ldots \   \ . \la{319}
 \ee
 Such redefinitions may be interpreted as quantum corrections to the integrability conditions \rf{condH}:
if  the original couplings  $\lh_p$ satisfied
 $f_a(\lh)=0$, then the corrected ones $\bar \lh_p$ would satisfy a \textit{corrected} version of the integrability conditions,
\be
\bar{f}_a(\bar \lh)   = 0\ , \qquad \bar{f_a} \equiv  f_a + \a'  \, Q_p \,  \del_{\lh_p} f_a   + \ldots \ \ .
\ee

\subsection{Some special  RG-stable  cases}      
There are still special exceptional cases of the integrable $G \!\times\!G/H$ model \rf{couH},\rf{bb1},\rf{condH}  that are automatically stable under  the  2-loop  RG flow 
in the \GB  scheme. Two of them are discussed   below. 

\subsubsection{$G\times G/H$ model related to $G/H$ $\l$-model}
One solution of the integrability conditions \rf{condH} is
 \be \la{3221}
 r=k\ , \  \ \ \rho_{21}=0 \ , \ \  \rho_{11} = \rho_{22} = k \ , \qquad {\rm i.e.}  \quad \ \  r=s = u = k\ ,\ \ \  t=b \ ,  \ee
 on which \rf{couH},\rf{bb1} become (redefining $(g,\wt g) \equiv (g^{(1)}, ({g^{(2)}})^{-1})$) 
\be \la{cl} \begin{aligned}
&\L =   -k' \Big( \L_{_{\rm WZW}}(g) + \L_{_{\rm WZW}}(\wt g) - \Tr \Big[ J_+ (g)  \big(P_H + \l' \,  P_{G/H}\big) K_-( \wt g)\Big] \Big)   \ ,  
 \\
&J_+(g)  \equiv g^{-1}\del_+ g \ , \qquad  K_- (\wt g) \equiv \del_- \wt g\,  \wt g^{-1} \ , \qquad \quad  
k' \equiv  -k \ , \quad \ \   \l' \equiv  k'^{-1}  t \ . 
\end{aligned} \ee
This model is a "gauged" version of \rf{212}, similarly being constructed from  a  combination of two WZW  Lagrangians  coupled  by a  current-current term.
This particular $G\times G/H$ model appears from the "tripled" formulation \cite{HLT2} of the $G/H$\  $\l$-model \cite{Sfetsos:2013wia}  (after removing a decoupled  third WZW  part). 
Compared to generic $G\times G/H$ models, the $G \times G$ global symmetry is enhanced to a chiral gauge symmetry $G(\xi^-) \times G(\xi^+)$ acting as (see footnote \ref{chir})
\be
(g, \, \wt g) \to \big(u(\xi^-) \, g,  \ \wt g \, v(\xi^+)\big) \ , \qquad \qquad \big(u(\xi^-), \, v(\xi^+)\big)\in G(\xi^-) \times G(\xi^+) \ .
\ee
This symmetry protects the structure of \rf{cl}, allowing only the coupling $\l'$ to run. 
The 1- and 2-loop $\beta$-functions in \rf{1H}--\rf{fH} and 
Appendix  \ref{GGHapp} 
lead to the following RG equation for $\l'$
\be \begin{aligned}
\ddt \l' = \frac{\cG \l'}{k'} + \frac{\cG \l'\big[c_{_H}-(2\cG-c_{_H})\l'^2\big]}{k'^2(1-\l'^2)}\ .
\end{aligned} \ee
This is the 2-loop $\beta$-function \cite{HLT2}
 for the $\l$-model based on the symmetric space $G/H$ with  parameters $(k,\l)$ related to $(k',\l')$ by  $k' = k+2\cG$, $\l' = \tfrac{k}{k+2\cG}(\l^{-1} + 2\cG)$.

\subsubsection{Integrable deformation of GMM model on $G\times G/H$   and  $T^{1,1}$ model}

Let us consider a particular solution  of the integrability conditions \rf{condH} that was studied in \cite{ABL}, 
\be \la{3244}
r=k\ , \qquad \qquad \r_{12}=\r_{21}=0  \ , \ \ \ {\rm i.e.} \ \  t=b=0 \ . \ee 
The Lagrangian of the corresponding theory \rf{couH},\rf{bb1} is given by 
\begin{align}
\L ={}&-\ha \, \Tr \big[ \rh \, P_+ P_-  + \wt{\rh} \,  \widetilde P_+ \widetilde P_- \big] -\ha k \, \Tr \big[ \B_+ \B_-  + \widetilde{\B}_+ \widetilde{\B}_-  -2 \B_+ \widetilde \B_- \big]  + k \, \big[ \L_{_{\rm WZ}} (g)  - \L_{_{\rm WZ}} (\widetilde g) \big]  \la{3c}   \ , 
\end{align}
where  we  have  set\foot{Ref. \cite{ABL}  used the notation $(k, \, \rh, \, \wt \rh) \equiv (\l^2, \, \l_2^2,  \, \l_1^2 )$.}
\be\la{3255}
  (g^{(1)}, g^{(2)})\equiv (g,\wt g)\ , \quad 
  \wt P_\pm =  P_\pm (\wt g)\ , \quad  \wt \B_\pm =  \B_\pm (\wt g)\ , \qquad \ \ \ \rh\equiv \rho_{11}=s\ , \quad  \wt \rh\equiv\r_{22}=u \ . \ee 
This is an integrable deformation of the special point $\rh=\wt \rh = k$ that corresponds to the conformal GMM model \ci{GMM} on the homogeneous space $G\times G/H$ with equal levels. 

 Specializing the 1-loop   and 2-loop $\beta$-functions in \rf{1H}--\rf{fH} and   Appendix \ref{GGHapp} to this case, 
 we find that the model \rf{3c} is automatically stable  under 2-loop renormalization 
  with only $\rh$ and $\wt \rh$ running,
\be \la{bg} \begin{aligned}
& \ddt \rh = 2 \cgg \big(1- \tfrac{k}{ \rh}\big) \Big( 1 + \tfrac{ 1}{ \rh} \Big[ 2(\cgg-\ch) -  (3\cgg-2\ch) \tfrac{k}{\rh}\Big] \Big)  \ , \qquad \\
&\ddt \wt \rh = 2 \cgg (1- \tfrac{k}{ \wt \rh}\big)  \Big( 1 + \tfrac{ 1}{ \wt \rh} \Big[ 2(\cgg-\ch) -  (3\cgg-2\ch) \tfrac{k}{\wt \rh}\Big]\Big)  \ , \quad \qquad
\ddt k =0 \ .
\end{aligned} \ee
Remarkably, the RG evolution of $\rh$ and $\wh$  is  decoupled.  Note that the  structure of their   $\beta$-functions  is  similar to the  one in the PCM$_k$ case   \rf{Pkb}. 
As expected, the GMM model $\rh=\wh=k$ is a fixed point. 

 Let us consider  the  simplest  example of this theory  \rf{3c}   with  $G= SU_2 $ and $H=U_1$  and   
choose the parametrization 
\be g=e^{\p_1 T_1} e^{\theta_1 T_2} e^{\psi T_3} \ , \qquad \qquad   \wt g=e^{-\p_2 T_1} e^{-\theta_2 T_2} e^{-\wt \psi T_3} 
 \ , \la{coo} \ee
 where the $SU_2$ generators are 
$T_A = \tfrac{i}{2} \s_A$   
 and  the generator of  $H=U_1$ 
  is  $T_3$. We shall fix the $H$ gauge freedom   by setting $\wt \psi=0$. As a result,   we get an integrable 
   5-dimensional \sm  (cf. \rf{13})
  \begin{align}
\L ={} & (G_{mn} + B_{mn})  \del_+ x^m \del_- x^n 
=\tfrac{1}{4} k \, \Big[ \del_+ \psi \del_- \psi + \cos^2{\th_1} \, \del_+\p_1 \del_- \p_1 + \cos^2{\th_2} \, \del_+\p_2 \del_- \p_2 \no \\
&\qquad +  2  \cos{\th_1}\,  \del_+ \p_1 \del_- \psi + 2   \cos{\th_2}\,  \del_+\psi \del_- \p_2 + 2 \cos{\th_1}\cos{\th_2} \,  \del_+ \p_1 \del_- \p_2 \Big] \no \\
&\qquad  + \tfrac{1}{4} \rh\,  \big[ \del_+ \th_1 \del_- \th_1 + \sin^2{\th_1}\,  \del_+ \p_1 \del_- \p_1 \big]+ \tfrac{1}{4} \wh \, \big[ \del_+ \th_2 \del_- \th_2 + \sin^2{\th_2} \, \del_+ \p_2 \del_- \p_2  \big] \ .   \la{mod}
  \end{align}
  The resulting  target space  geometry 
corresponds to 
 the  $T^{1,1}$  metric  
     and a  particular $B$-field \cite{ABL}\foot{Due to 
   differing conventions, the $B$-field here  is opposite in sign to that  in  \cite{ABL}.
   This difference is not significant, and can  be removed by a parity transformation.}
 \begin{align}
ds^2_{T^{1,1}} ={} & G_{mn} dx^m dx^n  = \tfrac{1}{4} k \,  (d\psi + \cos{\th_1} \, d\phi_1 + \cos{\th_2} \, d\phi_2)^2\no  \\
 &\qquad \qquad \qquad\ \ + \tfrac{1}{4} \rh \, (d \th_1^2 + \sin^2{\th_1} \, d \p_1^2)+ \tfrac{1}{4} \wh \,  ( d  \th_2^2 + \sin^2{\th_2} \, d \p_2^2 )  \ ,   \la{t11}\\
 B ={}& \ha B_{mn} dx^m  \wedge dx^n 
  = \tfrac{1}{4} k \, (d\psi + \cos{\th_1} \, d\phi_1) \wedge (d\psi + \cos{\th_2} \, d\phi_2) \ . \la{b11}
\end{align}
The 3 parameters $\rh,\wt \rh, k$ of \rf{3c} are thus mapped to the 3 parameters of the $T^{1,1}$ metric  in    \cite{romans}.\foot{\sloppy\la{f14}To recall, the $T^{1,1}$   metric 
\rf{t11} 
is
 an Einstein space 
 if $\rh=\wt \rh= {3\ov 2} k $. 
 It then serves  as a  base of a  Ricci flat 6d conifold  with metric $dr^2 + r^2 ds^2_{T^{1,1}}$ if we formally 
set 
$k=\tfrac{1}{9}$
 so that  $R_{ij} = 4 g_{ij}$. 
 In general, the non-zero components  of the  Ricci tensor of  the
 cone geometry 
 $ds^2= G_{mn}(X) dX^m dX^n= dr^2 + r^2  g_{ij}(x) dx^i dx^j$ (with $i=1, ..., d$) 
 are  ${R_{ij} (G)  = R_{ij} (g) -  (d-1) g_{ij}}$. Thus  it vanishes if $g_{ij}$ is an Einstein metric with a particular value of the scalar curvature 
 $R(g) = d (d-1)$ (this condition  is satisfied, e.g., for a unit-radius  sphere $S^d$  when  $G_{mn}$ is flat).  }
The 2-loop  RG equations  \rf{bg} become in this case ($ \cgg=2, \ \ch=0$)
\be \la{bet} \begin{aligned}
 \ddt \rh = 4 \big(1- \tfrac{k}{ \rh}\big) \big[ 1 + \tfrac{ 4}{ \rh} \big( 1 -  \tfrac{3\, k}{2\,  \rh}\big) \big]  \ , \qquad \qquad   \ 
&\ddt \wt h = 4 \big(1- \tfrac{k}{ \wt \rh}\big) \big[ 1 +  \tfrac{4 }{ \wt \rh} \big( 1 -  \tfrac{3\, k}{ 2\, \wt \rh}\big) \big]  \ .
\end{aligned} \ee
As we shall discuss in Section 4, the 2-loop RG stability of this  $T^{1,1}$ model  may be  understood 
 as  consequence 
of  the fact 
 that  the \sm \rf{mod} is self-dual  under T-duality  in the $\psi$-direction.


\section{Integrable $T^{1,q}$ model}

Let us  now  introduce a new  integrable \sm  with  
target space metric 
 $T^{1,q}$
  and a particular $B$-field,  which is a
  one-parameter generalization of  the $T^{1,1}$ model \rf{mod} of \cite{ABL}. 
  Its   special  conformal  case will be the $SU_2 \times SU_2/U_1$  GMM  model, 
  now 
  with \textit{unequal}  
   WZ levels \ci{GMM,PZT} (with their ratio related to
  the parameter  $q$). 

Our central observation is that,  starting with the $G \times G/H$ model \rf{couH}  and considering 
the case   when  the subgroup $H$ is {\it abelian},  the gauge invariance condition \rf{bb1} of \cite{ABL}    is too restrictive.
 At the particular point $\rho_{12}=\rho_{21}=0$, there is also a second 
 "branch" 
 of gauge invariant  models,\foot{Note that, in generic cases, WZ terms present a topological obstruction to gauging \cite{HS}. There are, however, special "anomaly-free" subgroups of the  WZ term's global symmetry $G_L\times G_R$ that can be gauged \cite{witten}, satisfying $\Tr_L[T_A T_B] - \Tr_R[T_A T_B] = 0$. This condition is satisfied here by the gauge transformations \rf{gI} and  \rf{42} on both "branches" of theories, due to cancellation between the two copies of $G$ in $G\times G/H$.}
\be
\begin{aligned}
&\L = -\ha \, \rho_{ij} \, \Tr[ P^{(i)}_+ P^{(j)}_- ] -\ha \, r_{ij} \, \Tr[ \B^{(i)}_+ \B^{(j)}_- ] + k_i  \, \L_{_{\rm WZ}}(g^{(i)}) \ ,
 \\
 &\rho_{12}=\rho_{21}=0 \ , \qquad r_{ij} = \begin{pmatrix} r  & q( - r - k_1) \\ q(-r +k_1) & q^2 r  \end{pmatrix} \  , \qquad q^2\equiv -k_2/k_1  \ , 
\end{aligned}
\la{bb2}
\ee
where $k_1$, $k_2$   are assumed to  be  of opposite sign. 
 The action  for \rf{bb2} 
 is  invariant under the modified  gauge transformation\foot{The reason for the 
  restriction of  $H$ to be abelian if $q\neq 1$ is that the  variation of the Lagrangian \rf{couH}  under \rf{42}  with 
 $\ww\in H$ will  be proportional to $(q-1)\L_{_{\rm WZ}}(\ww)$, which vanishes for abelian $H$   for any $q$. We also need to assume $\rho_{12} = \rho_{21}$ to prevent mixing between $P^{(1)}$ and $P^{(2)}$ terms, which transform differently  under $\ww$ and $\ww^q$ respectively.}
\be\la{42} 
(g^{(1)}, \, g^{(2)}) \to (g^{(1)} \ww^q , \,  g^{(2)} \ww) \ , \qquad \qquad  \ww=\ww(\xi^+, \xi^-)\in H \ .
\ee
Here $\ww^q$ is the  $q$-th power of the abelian  group element $\ww$.  
 In the case  when   the  abelian $H$  is compact
 then, 
to make   $\ww^q$   single-valued,  one  should  
 assume   that 
$q= \sqrt{ |{k_2\ov k_1}|}$ is an {integer}.\foot{More generally,  one could consider   a "twisted" action of the abelian subgroup, $(g^{(1)}, \, g^{(2)}) \to (g^{(1)} \ww^q , \,  g^{(2)} \ww^p)$  
characterized by integers $p,q$ satisfying $q^2/p^2=-k_2/k_1$.
In the $SU_2 \times SU_2/U_1$   example discussed below,   that would  lead to the  $T^{p,q}$  model.}
At the value $q=1$  (i.e. $k_1=-k_2$),  this model 
 intersects  with 
 the gauge invariant $G\times G/H$ model \rf{couH},\rf{bb1} considered above.
 
We 
claim 
that  model \rf{bb2} is integrable (admitting a Lax representation) 
 if\,\foot{The case $r=-k$ is also integrable since it is related to \rf{rk} by parity.}
 \be r=k  \ . \la{rk}
   \ee
 In this case it becomes a 
generalization of \rf{3c} to  
 the case of  
 unequal  levels $k,\wt k$,
\begin{align}
\hspace{-0.2cm}\L ={}&-\ha \, \Tr \big[ \rh \, P_+ P_-  + \wh \,  \widetilde P_+ \widetilde P_- \big]
-\ha  \, \Tr \big[ k\,  \B_+ \B_-  +  \wt k \, \widetilde{\B}_+ \widetilde{\B}_-  -2  \sqrt{k \wt k} \, \B_+ \widetilde \B_- \big]  + k \L_{_{\rm WZ}} (g)   - \wt k  \L_{_{\rm WZ}} (\widetilde g)  \  , \la{4c} \end{align}
where we have set (cf.  \rf{3255})
  \be \la{455}
    \begin{aligned}  &
    (g^{(1)}, g^{(2)})\equiv (g,\wt g)\ , \qquad  \wt P_\pm =  P_\pm (\wt g), \qquad   \wt \B_\pm =  \B_\pm 
  (\wt g),\ \\
& \rh\equiv \r_{11}, \qquad \wh\equiv \r_{22} \ , \qquad  k\equiv k_1, \quad \wt k \equiv - k_2 \ , \qquad 
q= \sqrt{ \tfrac{\wt k}{k}} \ .
 \end{aligned}\ee
 
 The fact that the $\wt k =k$ limit  \rf{3c}  is an integrable theory provides a first check of the integrability of \rf{4c}.
Indeed, 
starting from the Lax connections \cite{ABL} for \rf{3c} 
 (with $z$ as spectral parameter),\foot{These 
Lax connections 
were obtained  in \cite{ABL} from the affine Gaudin Lax connection of the general   integrable $G\times G/H$ model \rf{couH},\rf{bb1},\rf{condH}  by taking  the limit $r=k$, $\r_{12}=\r_{21}=0$. It was found that certain components of the Lax connection degenerate to zero and thus the flatness condition  of the resulting 
 connection $L_\pm$  does not imply 
 some of the equations of motion. 
  However, one can  consider a generalized limiting procedure 
   by  infinitely rescaling  the spectral parameter while taking this limit, thus 
   obtaining a second  connection $\wt L_\pm$ that "misses"  a different  subset of equations of motion. 
   The  flatness conditions of the two Lax connections   together  encode the full set of the equations
  of motion.
  The fact of having two separate Lax connections  may  seem   unusual but should be sufficient for the 
   integrability applications: for example, each Lax connection  will lead to its own family of conserved charges.
   Note also that for $k=0$  the  two Lax connections \rf{44},\rf{45}  become the  familiar ones 
    of the two decoupled  $G/H$ 
$\sigma$-models so the fact  of having two 
 connections may not be totally surprising   (we thank B. Hoare for this comment). 
   } 
\begin{align}
&L_+(z) = \B_+ + z^{-1} P_+ \ ,\qquad  \qquad L_- (z) = \frac{1}{k z^2 - \rh} \Big[ (k-\rh)(\B_- + z P_-) + k (z^2-1) \wt \B_- \Big] \ ,\la{44}\\
&\wt L_-(z) = \wt \B_- + z \wt P_- \ , \qquad \qquad \ \ \ 
\wt L_+(z) = \frac{1}{k z^{-2} - \wh} \Big[ (k-\wh)(\wt \B_+ + z^{-1} \wt P_+) + k (z^{-2}-1) \B_+ \Big]  \ ,\la{45}
\end{align}
we have found the following  Lax connections for \rf{4c} by replacing some   factors of $k$   by $\wt k$,
\begin{align}
&L_+(z) = \B_+ + z^{-1} P_+ \ , \la{Lf}
\qquad \qquad L_- (z) = \frac{1}{k z^2 - \rh} \Big[ (k-\rh)(\B_- + z P_-) +  \sqrt{k\wt k}(z^2-1) \wt \B_- \Big] \ ,\\
&\wt L_-(z) = \wt \B_- + z \wt P_- \ , \la{Ll} \qquad \qquad \ \ \
\wt L_+(z) = \frac{1}{\wt k z^{-2} - \wh} \Big[ (\wt k-\wh)(\wt \B_+ + z^{-1} \wt P_+) + \sqrt{k\wt k} (z^{-2}-1) \B_+ \Big]  \ .
\end{align}

 Assuming the  simplest case  $G=SU_2$, $H=U_1$ (see footnote \ref{an}), using the same  coordinate parametrization of this $SU_2 \times SU_2/U_1$ model as in \rf{coo}, 
and  fixing again the  $H=U_1$  gauge  as  $\wt \psi=0$,  we find the following   generalization of \rf{mod}
 \begin{align}
\L ={} & (G_{mn} + B_{mn})  \del_+ x^m \del_- x^n 
=\tfrac{1}{4} k \, \Big[ \del_+ \psi \del_- \psi + \cos^2{\th_1} \, \del_+\p_1 \del_- \p_1 + q^2 \cos^2{\th_2} \, \del_+\p_2 \del_- \p_2 \no \\
&\qquad +  2  \cos{\th_1}\,  \del_+ \p_1 \del_- \psi + 2q   \cos{\th_2}\,  \del_+\psi \del_- \p_2 + 2q \cos{\th_1}\cos{\th_2} \,  \del_+ \p_1 \del_- \p_2 \Big]  \la{modq} \\
&+ \tfrac{1}{4} \rh\,  \big[ \del_+ \th_1 \del_- \th_1 + \sin^2{\th_1}\,  \del_+ \p_1 \del_- \p_1 \big]+ \tfrac{1}{4} \wh \, \big[ \del_+ \th_2 \del_- \th_2 + \sin^2{\th_2} \, \del_+ \p_2 \del_- \p_2  \big] \ , \qquad  q=\sqrt{\tfrac{\wt k}{k}} \  . \no 
  \end{align}
The resulting  target space metric is that of the $T^{1,q}$  space \cite{romans}
 and the $B$-field is a natural generalization of the one in \rf{b11},
\begin{align}
ds^2_{T^{1,q}} ={} & G_{mn} dx^m dx^n  = \tfrac{1}{4} k \,  (d\psi + \cos{\th_1} \, d\phi_1 +q \cos{\th_2} \, d\phi_2)^2 \no
\\
 &\qquad \qquad \qquad
 + \tfrac{1}{4} \rh \, (d \th_1^2 + \sin^2{\th_1} \, d \p_1^2)+ \tfrac{1}{4} \wh \,  ( d \th_2^2 + \sin^2{\th_2} \, d \p_2^2 )  \ , \la{gq} \\
 B ={}& \ha B_{mn} dx^m  \wedge dx^n 
  = \tfrac{1}{4} k \, (d\psi + \cos{\th_1} \, d\phi_1) \wedge (d\psi + q\cos{\th_2} \, d\phi_2) \ . \la{bq}
\end{align}
 Like the $q=1$   case \cite{ABL}  in \rf{mod}, 
       the   presence of  the $B$-field  is  crucial here for   integrability
(the $T^{1,q}$ \sm  without $B$-field  is not  integrable \cite{bpz}).
The   coordinate form of the Lax connections \rf{Lf},\rf{Ll} is  ($T_A$ 
 are the $SU_2$ generators  in \rf{coo}) 
\begin{align}
&{L'}_+(z) = \cos{\theta_1} \, \del_+ \phi_1 \, T_3 + z^{-1} \big( \del_+ \th_1  \, T_2 + \sin \th_1 \, \del_+ \p_1  \, T_1 \big)\no \ , \\
&{L'}_- (z) = \tfrac{1}{k z^2 - \rh} \Big[ (k-\rh)\big( \cos \th_1  \, \del_- \p_1  \, T_3 + z  ( \del_- \th_1  \, T_2 + \sin \th_1 \, \del_- \p_1  \, T_1 )   \big) \no  \\
&\qquad \qquad \qquad \qquad \qquad 
 - (z^2-1) \big( \sqrt{k \wt k}   \, \cos \th_2 \, \del_- \p_2 + k \,  \del_- \psi \big)  \Big]    \ ,\\
&\wt L_+(z) = \tfrac{1}{\wt k z^{-2} - \wt h} \Big[ (\wt k-\wh)\big( - \cos \th_2  \, \del_- \p_2  \, T_3 + z^{-1}  ( - \del_+ \th_2  \, T_2 + \sin \th_3 \, \del_+ \p_3  \, T_1 )   \big) \no \\
&\qquad \qquad \qquad \qquad \qquad \  + \sqrt{k\wt k} (z^{-2}-1)\big( \cos \th_1 \, \del_+ \p_1 + \del_+ \psi   \big)\Big]  \ , \no\\
&\wt L_-(z) =  - \cos{\theta_2} \, \del_- \phi_2 \, T_3 + z \big( - \del_- \th_2  \, T_2 + \sin \th_2 \, \del_- \p_2  \, T_1 \big)  \ .
\end{align}
To simplify the expressions   we followed   \cite{ABL}  here in   replacing $L_\pm $ by its gauge transformed   version 
 ${L'}_\pm = \ww^{-1} L_\pm \ww + \ww^{-1} \del_\pm \ww$,   with $\ww=\exp (- \psi T_3)$. 

At the special point 
 $\rh=k, \ \wh=\wt k= q^2 k $,  the model  \rf{modq} 
 becomes the $SU_2\times SU_2/U_1$  case of the  conformal GMM   model  with   levels $k_1=k, \ k_2 =-\wt k$. 
 It was  pointed out in \cite{PZT} that the $SU_2\times SU_2/U_1$  GMM model corresponds to the 
 $T^{1,q}$ 
  metric  and 
  a particular $B$-field,   and its 2-loop conformality was explicitly  checked (see also \ci{belo}). 
  The  general GMM model has a  current algebra  symmetry \cite{GMM} and is  also  integrable in the    Lax connection   sense  \cite{BBS}.
 What we have shown above  is that 
  it   admits an  integrable extension \rf{modq} 
    away from the conformal point $\rh=k, \ \wh=\wt k$.

\subsection{Stability under the 2-loop RG flow}

Let us   now show   that the integrable  
$T^{1,q}$ model \rf{modq}   is  stable under the 2-loop RG flow. 

The general gauge invariant model \rf{bb2} (with the  $r=k$ 
condition   \rf{rk} relaxed) must be 
stable  under the RG   with  only $(r,\rh,\wt\rh)$  as running couplings.\foot{At the point $q=1$ or $k_1=-k_2$ where the two 
 "branches" 
 of gauge invariant theories \rf{bb2},\rf{bb1} intersect, one may worry that the couplings $\r_{12}$, $\r_{21}$ may also run, since this is no longer prevented by the gauge invariance.  However, in the abelian $H$ case 
 this is forbidden by an extra global "center" symmetry  \cite{romans, ABL},
$ (g,\, \widetilde g) \to (g\, z ,\, \widetilde g) \ , \ \   z\in 
{\rm Z}(H) 
 \subset H$ 
preserving the non-mixing of the   coset parts of the   current  $P$ and $\wt P$ in \rf{3c}. 
Note that  this symmetry alone would not be sufficient to explain the stability of the $T^{1,q}$ model since it does not prevent $r$ from running.} 
This is  due to  its $H$ gauge invariance and  global $G_L \times G_L$ symmetry prohibiting any  new  counterterm  structures. 
 We shall see that the $T^{1,q}$ model, obtained by fixing $r=k$, is a "fixed line" of its RG flow.

Relaxing  $r=k$  has the effect of replacing $k\to r$ in the metric, with $k$ still appearing in the $B$-field (cf. \rf{gq},\rf{bq})\foot{Rescaling $r\to  r' k  $  and $\psi\to { 1\ov \sqrt k   } \psi'$   this  background   can be put into the  form 
    symmetric under 
   $k\leftrightarrow \wt k, \ \rh \leftrightarrow  \wt \rh$:
   
   $\qquad  \ ds^2  = \tfrac{1}{4} r' \,  (d\psi' + \sqrt k \cos{\th_1} \, d\phi_1 + \sqrt {\wt k}  \cos{\th_2} \, d\phi_2)^2 + 
 \tfrac{1}{4} \rh \, (d \th_1^2 + \sin^2{\th_1} \, d \p_1^2)+ \tfrac{1}{4} \wh\,  (  d\th_2^2 + \sin^2{\th_2} \, d \p_2^2 )  \ , $
 \ \ 
 
 $
\qquad \ \ \    B = \tfrac{1}{4}  \, (d\psi' + \sqrt k \cos{\th_1} \, d\phi_1) \wedge (d\psi' + \sqrt {\wt k}  \cos{\th_2} \, d\phi_2) \ . $
   }
\begin{align}
ds^2  ={}& \tfrac{1}{4} r \,  (d\psi + \cos{\th_1} \, d\phi_1 + q \cos{\th_2} \, d\phi_2)^2 + 
 \tfrac{1}{4} \rh \, (d \th_1^2 + \sin^2{\th_1} \, d \p_1^2)+ \tfrac{1}{4} \wh\,  (  d\th_2^2 + \sin^2{\th_2} \, d \p_2^2 )  \ \no ,\\
  \la{bQ2} B ={}& \tfrac{1}{4} k \, (d\psi + \cos{\th_1} \, d\phi_1) \wedge (d\psi + q \cos{\th_2} \, d\phi_2)  \ , \qquad \ \  q= \sqrt {\tfrac{\wt k}{ k} }\ .  
\end{align}
   The corresponding 
      2-loop  $\beta$-functions in the \GB scheme   following from \rf{BMT}  are 
     ($k, \wt k$ do not run)\foot{Note that for $k=\wt k$ (i.e. $q=1$) this  system  of RG equations is   obviously  symmetric 
     under interchanging  $\rh$ and $\wt \rh$.
      Note also that 
     setting $k=\wt k \to  0 $  with ${k\ov \wt k}=1$   and $h=\rh$  the 1-loop $\beta$-functions become 
     $\ddt r = 4 {r^2 \ov  h^2}, \ \ddt h= 4 - 2{r\ov h}$  so that   $\ddt { r\ov h} =  6 {r\ov h^2 }
      (   {r\ov h} - { 2 \ov 3})$.  The  point ${r\ov h} = { 2 \ov 3}$   corresponds to the case when the $T^{1,1}$   metric  is 
an Einstein 
      space (cf. footnote \ref{f14}), 
      i.e. 
      $R_{mn}= \Lambda G_{mn}$  with $\Lambda = {16 \ov 9 r}$. }
  \begin{align}
&\ddt r = 2(r^2-k^2) \Big[  \tfrac{1}{\rh^2 }  + 
\tfrac{\wt k}{k} \tfrac{1}{{\wt \rh}^2 }  + 
  \tfrac{r^2-3 k^2}{ r } \big(\,  \tfrac{1}{\rh^4}  +   \tfrac{{\wt k}^2}{k^2} \tfrac{ 1}{\wt \rh^4}
  \big)\Big] \ ,\la{b3}\\
&\ddt \rh = 4\big[ 1 - \tfrac{r}{2\rh}(1+\tfrac{k^2}{r^2}) \big] 
+ \tfrac{2}{ \rh^3   r^2} \Big[   8 \rh^2 r^2     - 8 \rh k^2r  
  - 12  \rh r^3  + 4 r^2  k^2 +    3 k^4  + 5 r^4  \no  \\
 &\qquad \qquad\qquad \qquad \qquad \qquad \qquad \qquad  \qquad +   \tfrac{\wt k}{k}
  \tfrac{{\rh}^2}{\wt \rh^2} (r^2-k^2) (3r^2- k^2)   \Big] \ ,  \la{b4}  \\
&\ddt {\wt\rh} =  4\big[ 1- \tfrac{ \wt k}{k} \tfrac{ r }{{2\wt \rh  }}(1+\tfrac{k^2}{r^2})\big]  
 +\tfrac{2 }{  {\wt \rh}^3  r^2}\Big[8  {\wt \rh}^2 r^2 -8 {\wt \rh}  {\wt k}k r -12 \tfrac{\wt k}{k}  {\wt \rh} r^3 
  + 4 r^2 {\wt k}^2  
 +3 {\wt k}^2 k^2+5 \tfrac{{\wt k}^2}{k^2} r^4   \no\\
&\qquad \qquad\qquad \qquad \qquad \qquad \qquad \qquad  \qquad +
  \tfrac{\wt k}{k} \tfrac{{\wt \rh}^2}{\rh^2}  (r^2-k^2) (3r^2- k^2) \Big] \ . \la{b5}
\end{align}
Thus   
$r=\pm k$  are fixed lines 
 of 
 \rf{b3}, 
both at 1-loop and 2-loop order. 
 The  couplings $(r,\rh,{\wt \rh})$ grow   linearly with $\tau\to \infty $ in the UV (reflecting asymptotic  freedom), 
 while they decrease to the GMM fixed point $(r,\rh,{\wt \rh}) = (k,k,\wt k)$ in the IR.\foot{As was argued in \cite{GMM}, the GMM model on $G\times G'/H$ is an exact CFT, 
assuming at least one of  the  cosets  $G/H$ or $G'/H$ is a symmetric space (as is indeed the case for the 
 $SU_2 \times SU_2/U_1$ model).}

Specialising to the fixed line $r=k$  of \rf{b3},  
 the expressions \rf{b4},\rf{b5}  simplify, giving  the 
 2-loop  $\beta$-functions  of  the  integrable $T^{1,q}$ model \rf{4c},\rf{modq},
\begin{align}
& 
\ddt \rh = 4 \big(1- \tfrac{k}{ \rh}\big) \Big[ 1 + \tfrac{ 4}{ \rh} \big( 1 -  \tfrac{3\,k}{ 2\,\rh}\big) \Big]  \ ,  
 \qquad \qquad \qquad
 \ddt \wt h = 4 \big(1- \tfrac{\wt k}{ \wt \rh}\big) \Big[ 1 +  \tfrac{4 }{ \wt \rh} \big( 1 - \tfrac{3\, \wt k}{ 2\, \wt \rh}\big) \Big]  
\ .  \la{betq} 
\end{align}
These   are   a  natural generalization of the  $\beta$-functions  for the $T^{1,1}$ model     in \rf{bet} to the case of $\wt k \not= k$.
 Like  in \rf{bet}, the  RG evolution of $\rh$ and $\wh$  happens to be  decoupled (while
 this is  not the case for $r\not=k$
 in \rf{b4},\rf{b5}).

Let us  note that,  in addition to  the $r=k$  case of the $T^{1,q}$ model, 
  the $\s$-model   corresponding to    \rf{bQ2} admits  \textit{another} integrable limit,
 $\wt k = 0$. 
 In this case it  factorizes  into 
 a squashed $S^3$   with  WZ  term    and a round $S^2$.\foot{The   $\beta$-function \rf{b5} for  the  coefficient $\wt \rh$ then matches that of  the $S^2$
$\s$-model, 
 i.e.\ a special   case of the $G/H$
 symmetric space  
  $\beta$-function  \cite{HLT2}  
 (here $G/H = SO(3)/SO(2)$, i.e. $\cgg=2, \ \ch=0$): 
 
 \qquad $\ddt \wt h = 2\cgg + 4\cgg(\cgg-\ch) \wt h^{-1}  = 4 + 16 \wt h^{-1}$.}
  Then
   the $\b$-functions
   \rf{b3} and \rf{b4}
both 
   become the same
   as those of
 this 
   squashed $S^3$  model  
 in
 \ci{shifman} (for 1-loop $\beta$-functions see   \ci{yosh}).\foot{ The relation to the 
   notation used in  \ci{shifman} is 
   $\eta= k \rh^{-1}, \ \lambda^2= {2\pi}  \rh ^{-1} $, \ $ \kappa= 1 - r \rh^{-1} $. The fact that $r=k$ is a fixed line of \rf{b3} is consistent with the findings there in the $\wt k=0$ limit. 
   }
 Taking  further limits, the $\beta$-functions   \rf{b3},\rf{b4},\rf{b5} agree with  
 other  previously known expressions:
\begin{enumerate}[(i)]
\item   Setting  $\wt k=0$   
     and $r=\rh$, we get from \rf{bQ2}    
       the direct  sum of the  PCM$_k$  (round $S^3$ with a WZ term) 
 and the $S^2$  $\s$-model.  In this  case \rf{b3},\rf{b4} are  indeed equivalent to the $\b$-function 
 of PCM$_k$, i.e.  \rf{Pkb}  with $\cgg=2$. 
 
\item Setting  $\wt k = 0$ (i.e.\  $q=0$) 
   and  then   $k=0$, we 
   instead 
   get the  direct sum of a squashed $S^3$ (with no WZ term) and a round $S^2$.
    The  $\beta$-functions  for $r$ and $\rh$ agree with those of the   "squashed" PCM in  \cite{HLT2}  
  (with $G=SU_2$ 
   and the  "squashing" parameter $\varepsilon= {r\ov \rh}$):
    
 \noindent
 $
 \qquad \ddt r = {2r^2 }{\rh^{-2} }    +  {r^3}{\rh^{-4}}  \ , \qquad
 \ddt \rh = 4\big( 1 - \tfrac{1}{2} r \rh^{-1} \big) + {2}{ \rh^{-3 } } \big(   8 \rh^2      
  - 12  \rh r   + 5 r^2 \big)  \ .
$ 

\end{enumerate}
        
\subsection{Covariance under T-duality \la{T}}
One can argue that  the RG stability of the  integrable  $T^{1,q}$ model \rf{modq}, i.e.\ the presence of the fixed line $r=k$ of \rf{b3},   is related  to
its property of being  self-dual under T-duality in the isometric $\psi$-direction.
To see this, let us write the  Lagrangian \rf{modq}  
in the following form\foot{Note that  $ \widehat{\L}$
   becomes   simply quadratic in the fields 
 at the  GMM point $\rh=k,\  \wh= \wt k= q^2  k$. }
\begin{align}
&\L = \tfrac{1}{4} k \, \Big[ (\del_+ \psi +\ta_+) ( \del_- \psi + \tb_- ) - \ha \ta_+ \tb_-\Big]  + \widehat{\L}  \ , \la{413} \\
&\qquad \qquad  \ta_\pm \equiv   2\cos{\th_1}\, \del_\pm \phi_1\ , \qquad  \qquad   \tb_\pm\equiv   2 q \cos{\th_2}\, \del_\pm \phi_2\ ,   \la{notU}\\
&\widehat{\L} 
\equiv   
 \tfrac{1}{4}\rh  \Big[   \del_+ \th_1 \del_- \th_1 + (  \sin^2{\th_1}  + \tfrac{k }{\rh}  \cos^2{\th_1} )   \del_+ \p_1 \del_- \p_1 \Big]  
\no\\ &\qquad 
+  \tfrac{1}{4}  \wh  \Big[  \del_+ \th_2 \del_- \th_2 +  (  \sin^2{\th_2}   
+ \tfrac{k q^2}{\wh}  \cos^2{\th_2} )\, \del_+\p_2 \del_- \p_2 \Big]  \la{mmm} \ . 
  \end{align}
\sloppy
Starting from the interpolating Lagrangian (obtained by $\del_\pm \psi \to A_\pm$   and adding the condition ${\del_+ A_-  - \del_- A_+=0}$  with a  Lagrange multiplier $\bpsi$) 
\be
\L_{\rm int} =  \tfrac{1}{4} k \Big[ (A_+ + \ta_+) (A_- + \tb_- ) - \ha \ta_+ \tb_- - {\bpsi}(\del_+ A_-  - \del_- A_+) \Big] +  \widehat{\L}   \ , \la{int}
\ee
 and integrating out $A_\pm$, we obtain the  following  T-dual  Lagrangian 
\be\la{417}
\bar{\L} = \tfrac{1}{4}  k\Big[ (\del_+ \bpsi + \ta_+) ( \del_- \bpsi - \tb_- ) + \ha \ta_+ \tb_-\Big] + \widehat{\L} \ .
\ee
This  is the same  as
the original theory 
 \rf{413}, 
with $\psi \to \bpsi$ and a coordinate redefinition $\phi_2 \to -\phi_2$ (under which $\tb_- \to -\tb_-$).

To appreciate  the special  structure of 
\rf{413},
   let us  relax the condition $r=k$ and go back to  the 
 general  model \rf{bb2} 
  corresponding to the background \rf{bQ2}.
 Using again the notation \rf{notU}, 
  we find the following generalization of \rf{413}
 \begin{align}
\L = \tfrac{r+k}{8} \Big[ (\del_+ \psi +\ta_+) ( \del_- \psi + \tb_- ) - \ha \ta_+ \tb_-\Big]   + \tfrac{r-k}{8} \Big[ (\del_- \psi +\ta_-) ( \del_+ \psi + \tb_+ ) - \ha \ta_- \tb_+\Big]   + \widehat \L \ .  \la{m1}
\end{align}
Applying   the T-duality  $\psi\to \bpsi$  to \rf{m1} we get, instead of \rf{417},
\begin{align} 
\bar \L ={} & \tfrac{1}{4} r \Big[ \Big (\del_+ \bpsi +\tfrac{r+k}{2r}\ta_+ + \tfrac{r-k}{2r}\tb_+ \Big) \Big(\del_- \bpsi - \tfrac{r-k}{2r}\ta_-  - \tfrac{r+k}{2r}\tb_- \Big)
 + \tfrac{r+k}{4r}\ta_+ \tb_- + \tfrac{r-k}{4r} \ta_- \tb_+ \Big] +  \widehat \L   \la{m2} \ .  
%
\end{align}
For  general values of $r$  and  $k$,  
\rf{m2} is different from  \rf{m1}; the only self-dual theory where  \rf{m1} and \rf{m2} coincide is the $T^{1,q}$ model  \rf{413} corresponding to $r=k$ (or its parity-conjugate $r=-k$). 

\sloppy
By the standard  path integral argument,  the  T-dual models \rf{m1},\rf{m2}   should be  quantum-equivalent.\foot{In general,  the 
  T-duality  transformation rules   may be 
   subject to quantum  $\a'$ corrections  \ci{Tseytlin:1991wr}
    that may be attributed to extra  finite counterterms  resulting  from 
     integration over  the auxiliary gauge field $A_\pm$ (see, e.g.,  \ci{HLT1}).  If
      the kinetic term of the  isometric coordinate  is non-trivial, i.e.\ 
      the
      term 
      quadratic in $A_\pm $ 
      is  $A_+ M(x) A_-$, then the leading  quantum correction to the effective Lagrangian 
      is represented by   the   term  $\Delta \L \sim \a' \, \del_+ \log M \, \del_- \log M$ (as well as a shift of the dilaton \ci{bu}). 
     In the case of \rf{m1} we have $M=1$ and thus  this  correction  is absent.}
Since the model  \rf{m1}  is stable
under the RG 
 due to its symmetries,  
 with  the 3  running couplings $r,\rh,\wt\rh$,  
its T-dual \rf{m2} must also be stable.
 Given that   the self-dual  points ${r=\pm k}$ are part of both RG-stable families 
 \rf{m1} and \rf{m2}, then they must also 
 remain in both families after the renormalization. Hence ${r=\pm k}$ must be   fixed lines  of the RG flow.
     This   was  indeed  confirmed above 
     by the explicit computation of the $\beta$-functions  leading to 
    \rf{b3}.

\section{Concluding remarks}

In this   paper we discussed  some new
 instances 
 of  a close connection    between 
the conditions of  integrability and  a  consistent restriction of the RG flow to a subspace of couplings. 

We have found the  2-loop  $\b$-functions of  the 6-parameter  $G \times G$ model \rf{par} 
and  have shown that  its integrability condition \rf{cond} is automatically preserved by the RG flow. 
 In \cite{today}, the 1-loop $\beta$-functions for this integrable 
   model were written in a universal form in terms of the twist function, revealing a hidden simplicity. 
   It would be  interesting to see if the complicated  expressions we have found 
   for the 2-loop $\b$-functions (see  Appendix \ref{GGapp}) 
simplify on the 
"integrable surface"
 once expressed in terms of the 
    twist function.\foot{One may  try to follow  the method of \cite{today} at the 2-loop order,   computing the Riemann tensor and then the 2-loop  $\beta$-function in terms of the twist function. 
    It   would  also be interesting   to investigate  the connection  
    to  the "doubled"  approach of   \cite{hassler}  which   studied the model \rf{par},\rf{cond}  with 
 additional  integrable $\eta$- or $\l$-deformation parameters   turned on.}
  
 We   also studied  the  6-parameter gauged $G\times G/H$  model \rf{couH},\rf{bb1},
 which is integrable   under the conditions \rf{condH}. 
 The latter  were  found
  to be stable under the 1-loop  RG flow  but, in general,  require   a certain  deformation 
 (i.e.\ the addition of  finite counterterms)  at the 2-loop level
to preserve  integrability. 
  It is possible   that 
 there  exists  an extended  target space formulation of the $G\times G/H$   model 
 in which  no additional 2-loop counterterms  are  needed 
 (as was demonstrated  for 
 the $\l$-model  
  examples  in \ci{HLT2}). 
 
We have found  that there are  still  some special cases   in which integrable $G\times G/H$  models
are automatically stable under the 2-loop RG flow. One simple example is  the $T^{1,1}$ model of \ci{ABL}. 
We also   constructed   a    new   class of integrable $G\times G/H$ models  \rf{4c} in the case  when the subgroup  $H$ 
is abelian (see \rf{bb2},\rf{rk},\rf{4c}). 
For $G=SU_2$ and $ H=U_1$,  this  led to 
 an integrable $T^{1,q}$ 
 model  generalizing  the  $T^{1,1}$ model, 
which   
we also found to be stable under the 2-loop  RG flow  
for any value of the parameter $q$. 
 This  model 
 may be interpreted as
an integrable  deformation of the 
conformal 
GMM model 
 with unequal levels  \ci{GMM}. 
Since the GMM model  admits a  $G\times G'/H$   generalization (with $G\neq G'$),
this raises   the question of whether  there is a larger class of integrable  $G\times G'/H$ 
models that flow to 
such 
conformal 
theories.\foot{One obvious possibility is to   consider   some analytic continuations, e.g., 
 take  $G'$  to be a different  real form of the complexification of $G$ (assuming the resulting \sm  
 couplings 
  $G,B$  remain real). 
  For example,   the  counterpart  of  the $SU_2\times SU_2/U_1$ model 
 would be   $SL_2(\mathbb{R})\times SU_2/U_1$.\la{an}}
 Another open question is whether the integrable $T^{1,q}$ model  admits a description in terms of affine Gaudin models (like the $T^{1,1}$ case) or if it is outside of that formalism.

Given a \sm with running   couplings, it can be   promoted to a conformal theory  (and thus embedded into string theory) 
 by adding two light-cone  directions $u$ and $v$, replacing  the RG "time" in  the coupling constants  by $u$   and adding a dilaton
 linear in $v$ \ci{Tseytlin:1992pq}. Fixing the  light-cone  gauge on $u$, one then gets  back the original \sm with 
 "local"  couplings depending on  2d time according to the RG equations. 
It would be interesting to study 
   whether  the connection   between  the classical  Lax integrability  of such local-coupling models   and the  RG  evolution of  couplings
   observed in  \cite{HLT3}     applies   also  to the models  discussed in this paper.

\section*{Acknowledgments}

We  are grateful  to  B. Hoare  for   many  useful discussions of related  questions  and comments on the draft. 
We also   thank   F. Hassler  and M. Shifman 
  for   interesting discussions. 
NL was supported by the EPSRC grant EP/N509486/1.
AAT was supported by the STFC grant ST/P000762/1.
\bigskip

\appendix

\section{Deriving the  integrability conditions for the $G^N$ model} \label{A}
\def\theequation{A.\arabic{equation}}
\setcounter{equation}{0}
It was shown in 
 \cite{DLMV}
 that the coupled model \rf{cou} is integrable for particular choices of the couplings $(\rho_{ij}, k_i)$ corresponding to realisations of  the affine Gaudin models. 
Here  we shall  try to demonstrate 
 the converse statement: these affine Gaudin models are the \textit{only} integrable  cases of the  coupled models \rf{cou}.

We will assume a natural ansatz \rf{Lan} for the Lax connection, valued in $\Lie(G)$
(here we  explicitly  indicate  the summation over $i=1, ..., N$)\foot{While \rf{a1} is the natural ansatz for the Lax connection arising from affine Gaudin models, it does degenerate at certain points in coupling space. For example, taking $\rho_{ij}$ to be diagonal (i.e.\ decoupled PCM$_k$ models), one instead requires a Lax connection valued in $\Lie(G)^N$. Thus it would also be interesting to consider other ansatze for the Lax connection.} 
\be\la{a1}
L_+ = \sum_i \a_{i}(z) \, J_+^{(i)} \ , \qquad\qquad  L_- = \sum_i \b_{i}(z) \, J_-^{(i)} \ ,
\ee
where $z$ is the spectral parameter. 
The curvature of this Lax connection takes the form
\be
F_{+-}(L) = \sum_i \Big( \b_{i}(1-\a_{i}) \  \del_+ J_-^{(i)} - \a_{i}(1-\b_{i}) \  \del_- J_+^{(i)} \Big) + \sum_{i\neq j} \a_{i} \b_{j} \ [J_+^{(i)},J_-^{(j)}] \ . \la{cur}
\ee
The equations of motion of the  model \rf{cou}  are 
(for $G^N$ with arbitrary $N$)
\begin{align}
E_i &\equiv \sum_j \Big(  (\rho_{ij} - \delta_{ij} k_j) \ \del_+ J_-^{(j)} + (\rho_{ij} + \delta_{ij} k_j) \ \del_- J_+^{(j)} +  \rho_{ij} \ [J_+^{(i)},J_-^{(j)}] + \rho_{ji} \ [J_-^{(i)},J_+^{(j)}] \Big)=0 \la{eom} \ .
\end{align}
We note that \rf{cur} and \rf{eom} are the \textit{unique} ways to write these expressions without any terms of the form $[J_+^{(i)},J_-^{(i)}]$, which have been eliminated using the identity $F_{+-}(J^{(i)}) = 0$. 

If  the  model \rf{cou}  is integrable then,\foot{Here we are assuming integrability and deriving necessary conditions on the couplings. Thus we do not need to worry about whether the $v^i(z)$ in \rf{a4} are independent functions (which would be relevant for the converse question).} 
 for some 
  $v^i(z)$,
  we have
\be
F_{+-}(L) = \sum_i v^i(z) \, E_i \ , \la{a4} \ee
which implies that 
\be \la{eqs} \begin{aligned}
&\b_{j}(1-\a_{j}) = \sum_{i} v^i (\rho_{ij} - \delta_{ij} k_j) \ ,\qquad \qquad    \a_{j}(1-\b_{j}) = \sum_{i} v^i (- \rho_{ji} - \delta_{ij} k_j) \ , \\ &\qquad   \a_i \b_j  =  (v^i - v^j) \rho_{ij}  \ , \ \ \ i\neq j\  \  (\text{no summation}) \ . 
\end{aligned}
\ee
This is a system of $N + N + (N^2-N) = N^2+N$ equations. Fixing the freedom to redefine the spectral parameter by setting $v^1 = z$, there are $N+N+(N-1) =3N -1$ "artificial" variables, $\a_i, \b_i, 
v^{i\neq 1}$. 
 After solving for these, there are $(N^2+N)-(3N-1)=N^2-2N+1$ remaining equations to be solved for 
 the $N^2+N$ variables $\rho_{ij}, k_i$. After solving all the equations, this leaves $(N^2+N)-(N^2-2N+1)=3N-1$ free parameters for the integrable theory 
 (including the WZ levels, which may be continuous for non-compact groups).\foot{One might worry that the integrability constraints on the couplings $\rho_{ij}, k_i$  resulting from \rf{a4} might depend on the spectral parameter $v^1 = z$. However, this will not happen because there is a rescaling ambiguity  $E_i \to c_i E_i$ in the definition of the equations of motion \rf{eom}. One may thus rescale $E_1$ to effectively set $v^1=z=1$ in \rf{a4}. Since the constraints on the couplings from \rf{a4} must be invariant under such rescalings, then they must not depend on $z$.}

 Thus the space of integrable models is $(3N-1)$-dimensional,
 which 
 coincides with the number of free parameters 
  following from the affine Gaudin  construction  (see \cite{today} and refs.\ therein).

Specializing to the $N=2$ case of $G\times G$, this counting suggests a $5$-dimensional space of integrable models.  Then  the 6   free parameters  $(s,t,u,b ,k_1,k_2)$ in \rf{par}
should   be subject to  only one  relation to ensure  integrability. 
 Solving the equations \rf{eqs} in this case, one indeed obtains the condition \rf{cond}  
 originally found  from the affine Gaudin  construction. 

To summarize,  for general $N$,  the space of integrable models has the same dimension as the space of affine Gaudin models. It remains to understand if there may still be  extra branches 
of integrable theories  
 not corresponding to the affine Gaudin models (cf.\ the $G\times G/H$ models,  where this seems to be the case for abelian $H$, see Section 4).  For the $N=2$ case of $G\times G$ models, we found  exact matching between the space of  integrable  models \rf{eqs} and the space of affine Gaudin models satisfying  the condition \rf{cond}.

\section{Explicit form of the  2-loop $\beta$-functions  \la{B}} 
\def\theequation{B.\arabic{equation}}
\setcounter{equation}{0}

Here  we shall provide  the explicit formulae for the  2-loop $\beta$-functions  of the  general 
$ G\times G$ and $G\times G/H$   models 
that were
used in 
 the main text.\foot{The formulae derived in this Appendix are also available in the Mathematica file attached to the arXiv submission of this paper.} We will also briefly explain how they were derived.

\subsection{$G\times G$ model \la{GGapp}}
For the $G\times G$ model \rf{par}, let us use the notation
\be \la{not}
\rho_{ij} = \rh_{(ij)} + b_{[ij]}
= \begin{pmatrix}
s & t \\ t& u
\end{pmatrix}
+ \begin{pmatrix}
0& b\\ -b& 0
\end{pmatrix} \ .
\ee
Let the  $2\times 2$ matrix $n_{ij}$ be the 
 "square root" of $\rh_{ij}=\rh_{(ij)}$, and let $m_{ij}$ be its inverse,
\be
n_{ik}\, n_{jk} = \rh_{ij} \ , \qquad m_{ik} \, n_{kj} = \delta_{ij} \ , \qquad \qquad n_{ij} = n_{(ij)}, \quad \ m_{ij} = m_{(ij)} \ . \la{not2}
\ee
The target space metric of the \sm \rf{par} is "diagonalized" by the vielbein
1-form
\unskip\foot{We use  the generators $T_A$ satisfying $[T_A, T_B] = i {f^C}_{AB} T_C$ and we define $f_{ABC} = \Tr[T_C T_D] {f^D}_{AB}$. For simple groups $G$,  the structure constants satisfy 
 ${f^A}_{BC} {f^B}_{AD} = -2 \cG \Tr[T_C T_D]$ 
 and ${f^D}_{EA} {f^E}_{GB} {f^G}_{DC} = \ccg f_{ABC}$, where $\ccg$ is the dual Coxeter number of $G$. Note that $ i {f^A}_{BC}$ in our present 
conventions   is equivalent to  ${f^A}_{BC}$ in the conventions of \cite{HLT2}.}
\be \begin{aligned}
&E^{Ai} = n_{ik} \, J^{(k)A} 
\ , \qquad A=1,\ldots,\dim{G} \ , \ \ i = 1,2 \ ,  \la{viel} \\
&J^{(k)} \equiv T_A J^{(k)A}=   \big( g^{(k)} \big)^{-1}  d g^{(k)} \ . 
\end{aligned} \ee
Then  the  coefficients of  the metric $ds^2 = G_{Ai, Bj} E^{Ai} E^{Bj}$ and the   3-form   $H = d B =  \tfrac{1}{6} H_{Ai, Bj, Ck} E^{Ai} \wedge E^{Bj} \wedge E^{Ck}$ are given by\foot{The overall factor of $i$ in  \rf{Hf} simply reflects the fact  that the vielbein \rf{viel} is imaginary. This makes no difference and could be eliminated by just multiplying $E^{Ai} \to i E^{Ai}$.}
\begin{align}
&G_{Ai, Bj} = -\ha \, \Tr[T_A T_B] \, \delta_{ij} \  , \la{Gf}\\
&H_{Ai, Bj, Ck} = \tfrac{i}{2} f_{ABC} \big[ k_l  \, m_{il} \, m_{jl} \, m_{kl} + b_{lp}(  m_{ip} \, m_{jl} \, m_{kl} +  m_{il} \, m_{jp} \, m_{kl} +  m_{il} \, m_{jl} \, m_{kp}) \big]  \ . \la{Hf}
\end{align}
From  Cartan's structure equation $d E^{Ai} + \widehat{\w}^{Ai}{}_{Bj} \wedge  E^{Bj} = T^{Ai}$ with torsion $T^{Ai} = \ha  {H^{Ai}}_{Bj, Ck} E^{Bj} \wedge E^{Ck}$, we obtain the torsionful spin connection
\begin{align}
&\widehat{\w}^{Ai}{}_{Bj} = \tfrac{i}{2} {f^A}_{BC} M_{ijk} E^{Ck}\ , \\
&M_{ijk} \equiv  m_{il}m_{jl}n_{kl} -  m_{il}n_{jl}m_{kl} -  n_{il}m_{jl}m_{kl}  \la{me}\\
&\qquad\quad  +  k_l  \, m_{il} \, m_{jl} \, m_{kl} + b_{lp}(  m_{ip} \, m_{jl} \, m_{kl} +  m_{il} \, m_{jp} \, m_{kl} +  m_{il} \, m_{jl} \, m_{kp})  \ . \no
\end{align} 
The torsionful Riemann curvature $\widehat{R}^{Ai}{}_{Bj} \equiv \ha \widehat{R}^{Ai}{}_{Bj,Ck,Dl}  E^{Ck} \wedge E^{Dl} = d \widehat{\w}^{Ai}{}_{Bj} + \widehat{\w}^{Ai}{}_{Ck} \wedge \widehat{\w}^{Ck}{}_{Bj} $ is then found in terms of $M_{ijk}$ to be
\be \begin{aligned}
\widehat{R}^{Ai}{}_{Bj,Ck,Dl} ={}& \tfrac{1}{4} \big[  2{f^A}_{BE} {f^E}_{CD} M_{ijp} n_{pq} m_{kq} m_{lq}  \\
&\qquad \qquad + {f^A}_{CE} {f^E}_{BD} M_{ipk}M_{pjl} - {f^A}_{DE} {f^E}_{BC} M_{ipl}M_{pjk} \big] \ . \la{Ri}
\end{aligned}\ee

It is then straightforward to substitute \rf{Hf},\rf{me},\rf{Ri} into the 2-loop $\beta$-functions in the \GB scheme \rf{BMT}, obtaining  explicit formulae   for 
 the RG equations    $\ddt \rho_{ij} = \b_{ij}(n_{11}, n_{12}, n_{22}, b, k_1, k_2)$ depending on the components of $n_{ij}$. Using a computer symbolic algebra package (e.g.\ Mathematica) it is easy to rewrite these expressions in terms of the components $s,t,u$ of the "square" coupling  $\rh_{ij} = n_{ik} n_{jk}$ in \rf{not}, with all  the square roots cancelling out
  as the Riemann tensor and the $H$-field must clearly be rational functions of $\rh_{ij}$. We thus obtain the $\beta$-functions
   in the  form  given  in \rf{2c},
\be \begin{aligned}
 & \qquad\qquad \ddt \rho_{ij}  = \a' \, \b_{ij}^{(1)} + \a'^2\,  \b_{ij}^{(2)} + \ldots \ , \\
 \b_{ij}^{(1)} = {}& \ccg (su-t^2)^{-2} \, F^{(4)}_{ij}(s,t,u,b ,k_1,k_2) \ ,  \qquad  
  \b_{ij}^{(2)} = {\rm c}_{_G}^2 (su-t^2)^{-5} \, F^{(9)}_{ij}(s,t,u,b ,k_1,k_2) \ , 
\end{aligned} \ee
where   the explicit form of the homogeneous polynomials  $F^{(4)}_{ij}$   and $F^{(9)}_{ij}$ is:
\ \\

\bigskip
\hspace{-1.26cm}\includegraphics[scale=0.37]{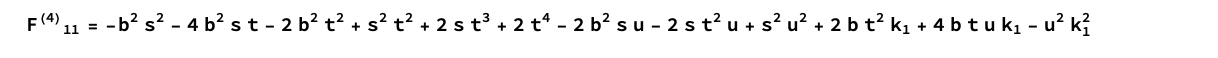}
\medskip

\hspace{-1.26cm}\includegraphics[scale=0.37]{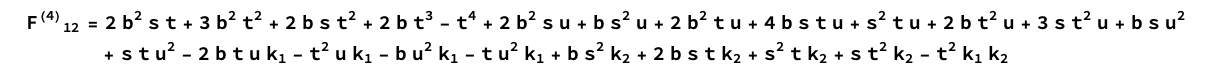}
\medskip

\hspace{-1.26cm}\includegraphics[scale=0.37]{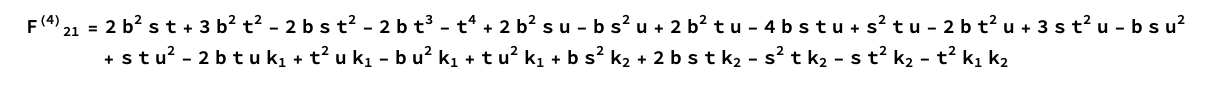}
\medskip

\hspace{-1.26cm}\includegraphics[scale=0.37]{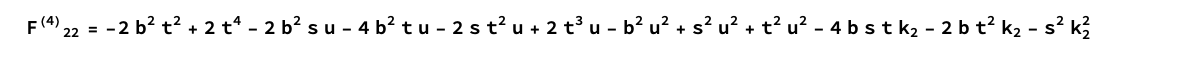}
\medskip

\hspace{-1.3cm}\includegraphics[scale=0.39]{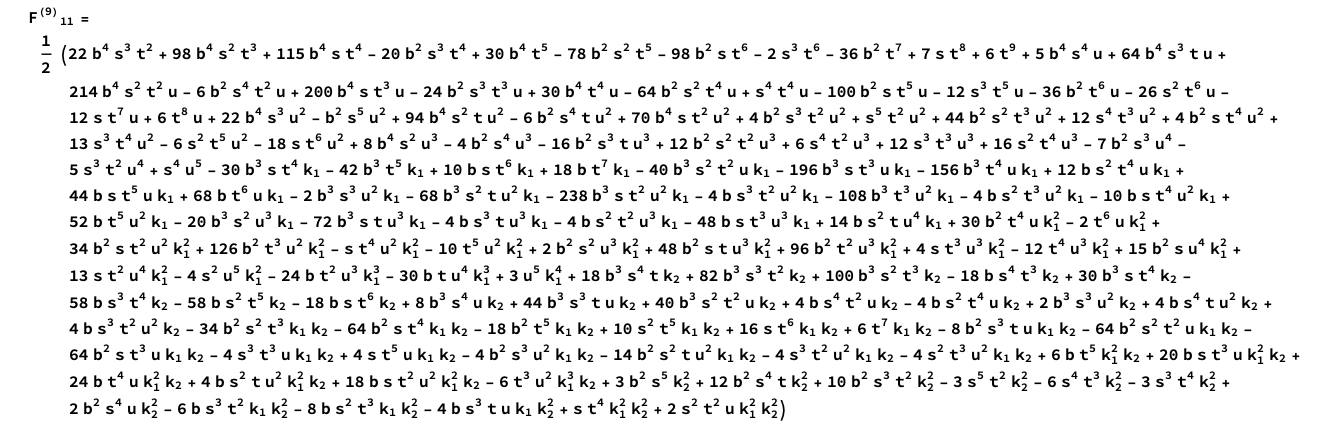}
\medskip

\hspace{-1.3cm}\includegraphics[scale=0.39]{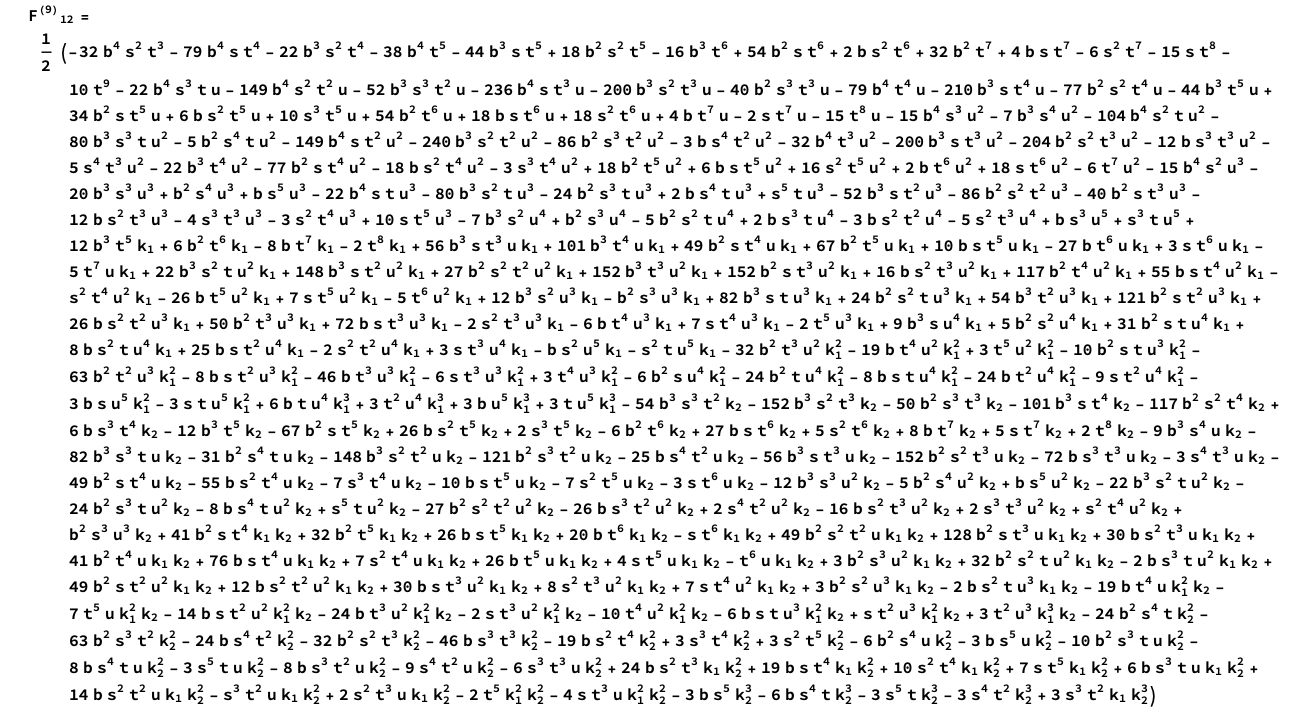}
\medskip

\hspace{-1.3cm}\includegraphics[scale=0.39]{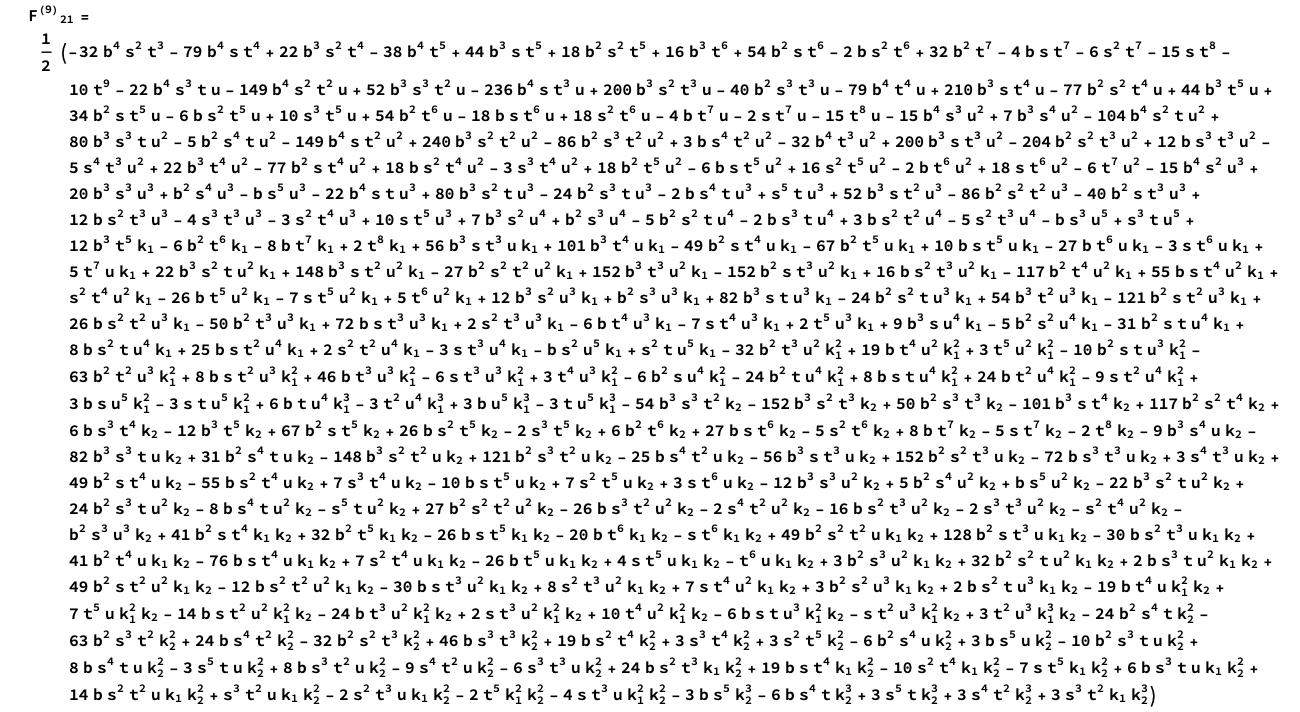}
\medskip

\hspace{-1.3cm}\includegraphics[scale=0.39]{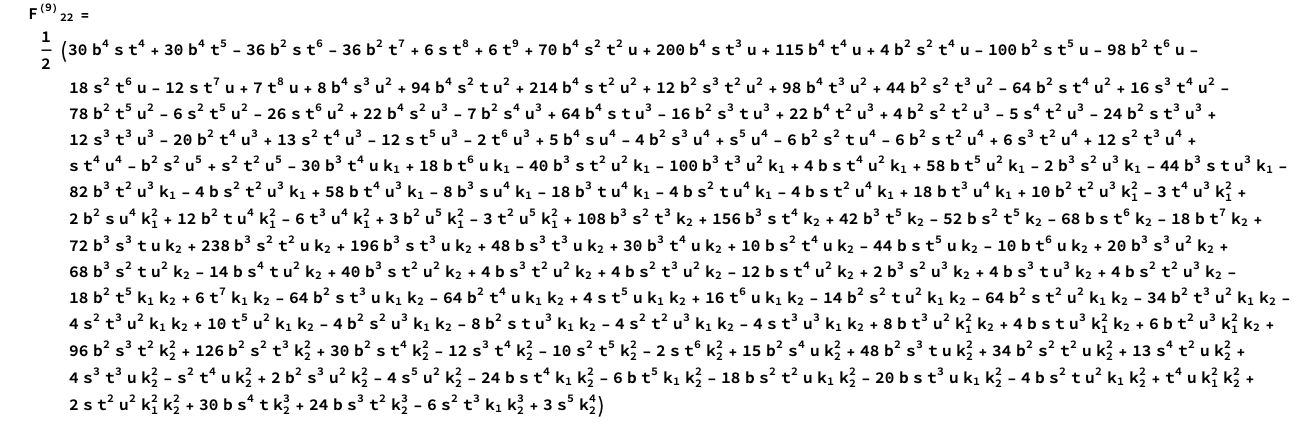}

\subsection{$G\times G/H$ model \la{GGHapp}}
The computation of the $\beta$-functions for the gauge invariant $G\times G/H$ model \rf{couH},\rf{bb1} is similar to the $G\times G$  case  above, except that one has to correctly handle the gauge invariance.

We shall again use the notation \rf{not} and \rf{not2},
  with  the symmetric "square root" of 
$\rh_{ij} = \rho_{(ij)}$
 being  $n_{ij}$, and its inverse being  $m_{ij}$.
In the computation below, we shall denote certain combinations of $n_{ij}$ and $m_{ij}$ by 
\be \begin{aligned}
&\hij  = m_{1i}m_{1j} + m_{2i}m_{2j} \ , \quad & &\nu_{ij} = n_{1i}m_{1j} - n_{2i} m_{2j}  \ , \\
&\Hij = \tfrac{k}{2} \hij  + \tfrac{b}{2}(m_{1i} m_{2j} + m_{2i} m_{1j}) \ , \quad &  &\l_{ij} =  m_{1i}m_{1j} - m_{2i}m_{2j} \ .
\end{aligned} \ee
We shall split  up the generators  $T_A$ of $G$ into $T_\a \in \Lie(H)$ and $T_a\in \Lie(G)/\Lie(H)$ (which are orthogonal with respect to the Killing form). 

Assuming from the beginning that the matrix $r_{ij}$ satisfies the gauge invariance condition \rf{bb1},\foot{Alternatively, one could obtain the same results by starting with $r_{ij}$ unconstrained, i.e.\ without gauge invariance imposed. One could  first  compute the torsionful Riemann tensor for the target space geometry \rf{couH} with general $r_{ij}$, $\rho_{ij}$. The gauge invariance condition \rf{bb1}  would 
 then  be  imposed and the resulting Riemann tensor   projected onto the non-degenerate directions of the metric $G_{MN}$.}  the target space metric of the \sm \rf{couH} is diagonalized by the
  vielbein\foot{The index  $M$   denotes  all tangent space directions.
 In the $G\times G$ case  in \rf{viel}  we had  $M= (Ai)$,   while here 
$
M = (\a, \bar \a, a i)$. Both $G$ and $H$ are assumed to be simple.
}
\be \begin{aligned}
&E^M = (e^\a, e^{\ol \a} , e^{ai}) = \Big(\sqrt{r}(\B^{(1)\a} -\B^{(2)\a}) \, ,\  \B^{(1)\a} +\B^{(2)\a}\, ,\  n_{ik} P^{(k)a} \Big) \ , \\
& \B^{(k)} \equiv  T_\a \B^{(k)\a} = P_H \big[ \big(g^{(k)}\big)^{-1} d g^{(k)} \big] \ , \qquad P^{(k)} \equiv  T_a P^{(k)a} = P_{G/H} \big[ \big(g^{(k)}\big)^{-1} d g^{(k)} \big]  \ .
\end{aligned}\ee
In this frame, the metric $ds^2 = G_{MN} E^{M} E^{N}$ and  the 3-form   $H = \tfrac{1}{6} H_{MNP} E^{M} \wedge E^{N} \wedge E^{P}$ have the following non-zero components
\begin{alignat}{3}
&G_{\a\b} = -\ha \, \Tr[T_\a T_\b ] \ , \qquad &&G_{ai, bj} = -\ha \, \Tr[T_a T_b] \  ,\\
&H_{\a\b\g} = \tfrac{i}{2} k r^{-3/2} f_{\a\b\g} \ , \qquad &&H_{\a,bi,cj} = i r^{-1/2} \Hij  f_{\a bc} \ . \la{HH}
\end{alignat}

The $H$ gauge invariance is reflected in the vanishing of all $\ol \a$ components of $G_{MN}$ and $H_{MNP}$, and,  in particular,  the fact that $G_{MN}$ is degenerate as a result. One could explicitly fix a gauge, eliminating some target space directions and removing this degeneracy. Instead, we find it more convenient to lift the degeneracy with a small parameter $\eps$ acting as a regulator,\foot{\sloppy
The use of the "regulator" $\eps$ is a short-cut for the following gauge-fixing procedure. Fixing an "axial" gauge ${i X^u (I^{(1)}_u + I^{(2)}_u) = y(\xi) \in \Lie{H}}$ ($u=1,2$  is the 2d index), the path integral  should be 
 independent of the choice of the constant 2d vector $X^u$ and the algebra-valued  function of 2d coordinates 
 $y(\xi)$. Inserting the $\delta$-function  of the gauge fixing condition into the path integral   and then 
   integrating over $X^u$ and $y$ with a Gaussian measure, i.e. 
 $ e ^{-\ha  X^u X_u  - \ha \eps \int d^2 \xi  \, \Tr[y y ] }$,  the result  should be  independent of $\eps$ (here we assume 
  Euclidean 2d signature but the  same  is true  also in Minkowski signature after an  analytic continuation). 
  Integrating  first over $y$ we get 
  $ \int d^2 X \exp \big[ {- \ha X^u X^v \big( \d_{uv} -  \eps \,   \Tr[(I^{(1)}_u + I^{(2)}_u) (I^{(1)}_v + I^{(2)}_v)] \big)}\big]$. 
  Integrating over $X_u$ restores the 2d Euclidean invariance  and  the  result  to leading order in the $\e\to 0$ limit 
  is equivalent to simply adding the regulator term $\Delta \L = -\ha \eps \, \Tr[ (I^{(1)}_u + I^{(2)}_u)^2]$  corresponding to  \rf{reg}.}
\be
G_{\ol \a \ol \b} = -\ha \eps \, \Tr[T_\a T_\b ] \ . \la{reg}
\ee
Computing the torsionful Riemann tensor as in  subsection 
 \ref{GGapp}, 
 one finds that it has a finite $\eps\to 0$ limit.
  This means that the resulting Riemann tensor for $\eps=0$ is unambiguous (since there are no divergences that could create finite-term ambiguities). Finally,  we project out the $\ol \a$ directions to obtain the non-zero components
\be \la{RR}
\begin{aligned}
&\widehat{R}^\a{}_{\b\d\e} = -\tfrac{1}{4r} f^\a{}_{\b\g} f^\g{}_{\d\e} + \tfrac{k^2}{4r^3} ( f^\a{}_{\d\g} f^\g{}_{\b\e}  - f^\a{}_{\e\g} f^\g{}_{\b\d}  ) \ , \\
&\widehat{R}^\a{}_{\b,dk,el} = (\tfrac{k}{2r} \l_{kl} -\ha p_{kl}) f^\a{}_{\b\g} f^\g{}_{de} + \tfrac{1}{r} A_{kj}A_{lj}( f^\a{}_{dc} f^c{}_{\b e}  - f^\a{}_{ec} f^c{}_{\b d}  ) \ ,\\
&\widehat{R}^\a{}_{b i,\d,el} = -\tfrac{1}{2r} A_{ji} g_{jl} f^\a{}_{bc} f^c{}_{\d e} - \tfrac{k}{2r^2} A_{li} f^\a{}_{\d \g} f^\g{}_{b e}  + \tfrac{1}{r} A_{lj}C_{ji} f^\a{}_{ec} f^c{}_{b \d}  \ , \\
&\widehat{R}^{ai}{}_{bj,dk,el} = (C_{ij} \l_{kl} -\ha \d_{ij} p_{kl}) f^a{}_{b\g} f^\g{}_{de} + \tfrac{1}{r} A_{ki} A_{lj} f^a{}_{d\g} f^\g{}_{b e}  -  \tfrac{1}{r} A_{li} A_{kj} f^a{}_{e\g} f^\g{}_{b d}    \ , \\
&\widehat{R}^{ai}{}_{bk,\d,\e} = - \tfrac{1}{4r} \d_{ik} f^a{}_{b\g} f^\g{}_{\d\e} + \tfrac{1}{r} C_{ij} C_{jk} (f^a{}_{\d c} f^c{}_{b \e}  -  f^a{}_{\e c} f^c{}_{b \d}  )  \ , \\
&\qquad A_{ij} \equiv \tfrac{1}{4}(\nu _{ji}-\nu _{ij}) - \chi_{ij} + \tfrac{r}{2} \l_{ij} \ ,  \qquad C_{ij} \equiv  -\tfrac{1}{4}(\nu_{ji}+\nu_{ij}) + \chi_{ij} + \tfrac{r}{2} \l_{ij} \ . \\
\end{aligned}
\ee

All that remains is to substitute \rf{HH},\rf{RR} into the 2-loop $\beta$-functions \rf{BMT} in the \GB scheme. The resulting expression is hard to evaluate as it contains hundreds of terms, each proportional to a contraction of the form $(f f f f)_{\a\b}$ or 
$(f f f f)_{ab}$ 
where each $f$ denotes a component $f_{\g\d\e}$ or $f_{\g d e}$ of the structure constants, and indices are contracted using the Killing form $\Tr[T_A T_B]$. One can show, however, that there are only 11 independent  such contractions after accounting for the antisymmetry of the structure constants.
This  allows for 
the efficient evaluation of the resulting $\beta$-functions for $r$ and $\rho_{ij}$. These are first obtained depending on $n_{11}, n_{12}, n_{22}$ but, as discussed in subsection \ref{GGapp}, they may be rewritten in terms of $s,t,u$ with all square roots cancelling.  As a result,     we obtain the 2-loop generalization of the 1-loop $\beta$-functions in \rf{1H}--\rf{fH} 
\be
\ddt \lh_p  = \a' \b_{\lh_p}^{(1)}+ \a'^2 \b_{\lh_p}^{(2)} \ , \qquad \qquad \lh_p\equiv (r,s,t,b,u) \ ,
\ee
where $\b_{\lh_p}^{(2)}$ are given by the following expressions:
\ \\

\bigskip
\hspace{-1.3cm}\includegraphics[scale=0.39]{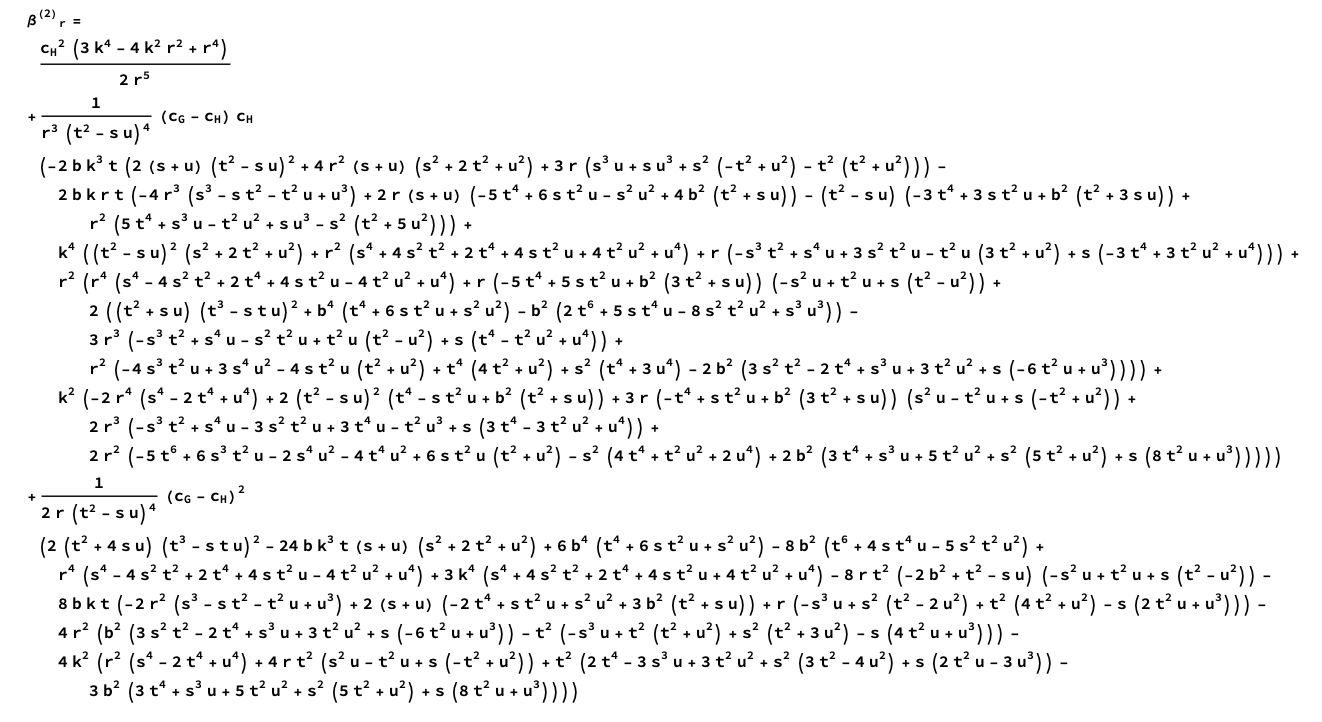}
\medskip

\hspace{-1.3cm}\includegraphics[scale=0.39]{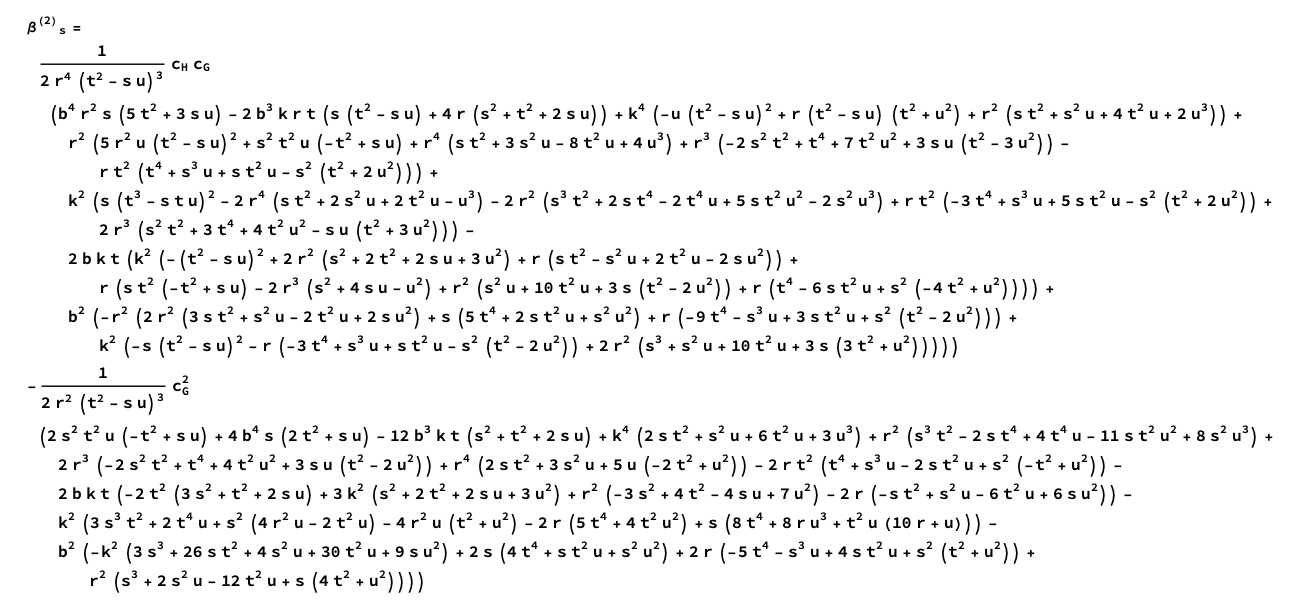}
\medskip

\hspace{-1.3cm}\includegraphics[scale=0.39]{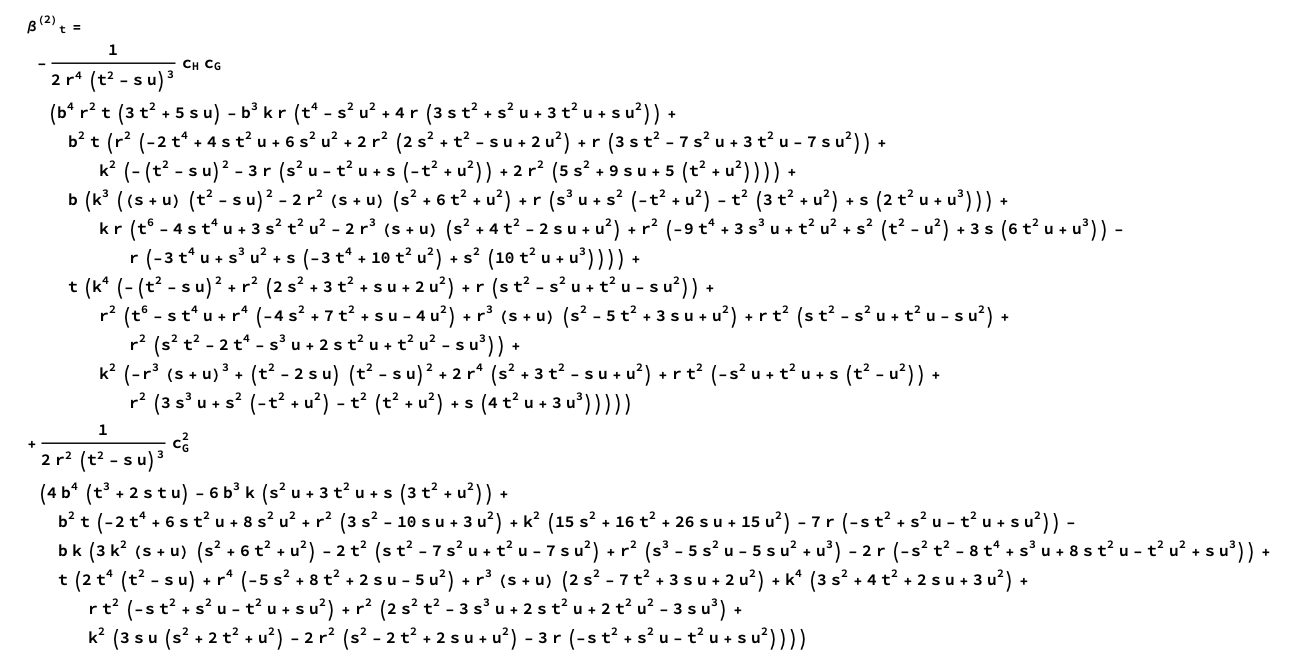}
\medskip

\hspace{-1.3cm}\includegraphics[scale=0.39]{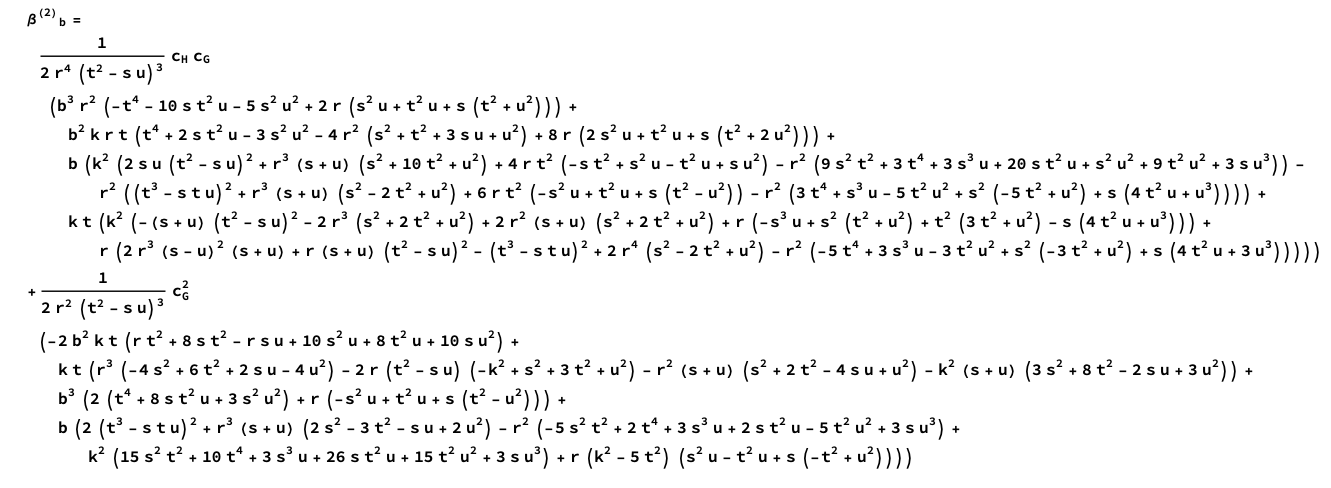}
\medskip

\hspace{-1.3cm}\includegraphics[scale=0.39]{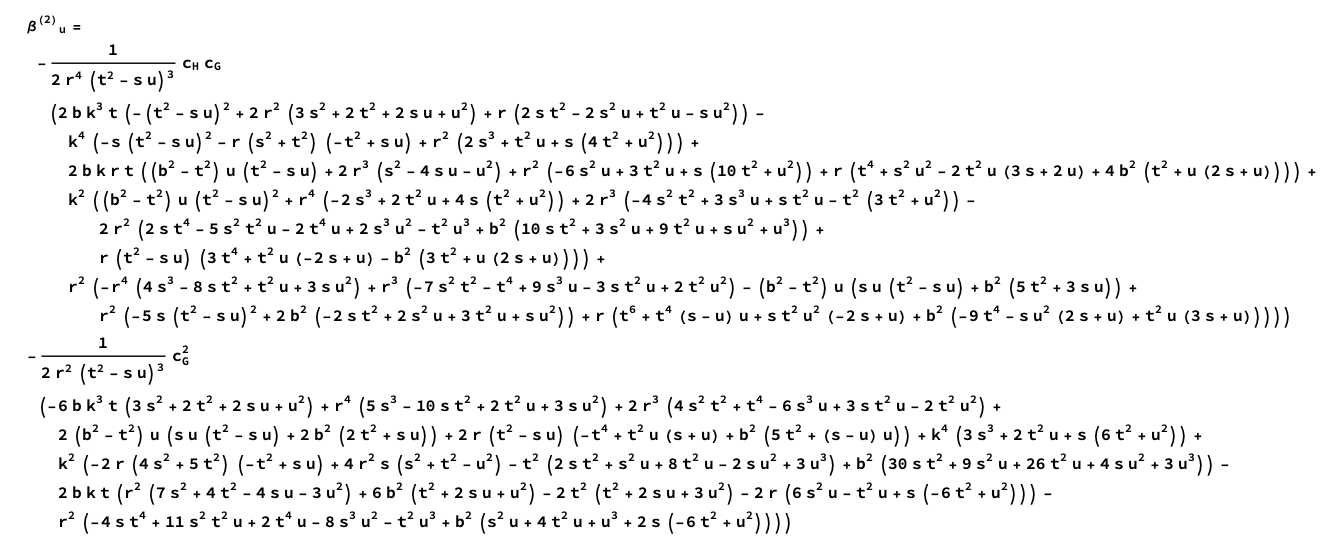}
\medskip

\small



\begin{thebibliography}{30}

\bibitem{intRG} 
V.~A.~Fateev, E.~Onofri and A.~B.~Zamolodchikov,
``Integrable deformations of the $O(3)$ sigma model. The sausage model,''
\doilink{Nucl.\ Phys.\ B {\bf 406}, 521 (1993)}{10.1016/0550-3213(93)90001-6}.

V.~A.~Fateev,
``Classical and Quantum Integrable Sigma Models. Ricci Flow, ``Nice Duality'' and Perturbed Rational Conformal Field Theories,''
\doilink{J. Exp. Theor. Phys. \textbf{129}, no.4, 566-590 (2019)}{10.1134/S1063776119100042}
\arxivlink{1902.02811}.

S.~L.~Lukyanov,
``The integrable harmonic map problem versus Ricci flow,''
\doilink{Nucl.\ Phys.\ B {\bf 865}, 308 (2012)}{10.1016/j.nuclphysb.2012.08.002}
\arxivlink{1205.3201}.
 
\bibitem{HLT1} 
B.~Hoare, N.~Levine and A.~A.~Tseytlin,
``Integrable 2d sigma models: quantum corrections to geometry from RG flow,''
\doilink{Nucl. Phys. B \textbf{949}, 114798 (2019)}{10.1016/j.nuclphysb.2019.114798}
\arxivlink{1907.04737}.

\bibitem{HLT2} 
B.~Hoare, N.~Levine and A.~A.~Tseytlin,
``Integrable sigma models and 2-loop RG flow,''
\doilink{JHEP {\bf 1912}, 146 (2019)}{doi:10.1007/JHEP12(2019)146}
\arxivlink{1910.00397}.

\bibitem{DLMV}
F.~Delduc, S.~Lacroix, M.~Magro and B.~Vicedo,
``Integrable Coupled $\sigma$ Models,''
\doilink{Phys. Rev. Lett. \textbf{122}, no.4, 041601 (2019)}{doi:10.1103/PhysRevLett.122.041601}
\arxivlink{1811.12316};
``Assembling integrable $\sigma$-models as affine Gaudin models,''
\doilink{JHEP \textbf{06}, 017 (2019)}{doi:10.1007/JHEP06(2019)017}
\arxivlink{1903.00368}.

\bibitem{ABL}
G.~Arutyunov, C.~Bassi and S.~Lacroix,
``New integrable coset sigma models,''
\doilink{JHEP {\bf 2103}, 062 (2021)}{doi:10.1007/JHEP03(2021)062}
\arxivlink{2010.05573}.
  
\bibitem{klim} 
C.~Klimcik,
``Yang-Baxter sigma models and dS/AdS T-duality,''
\doilink{JHEP \textbf{12}, 051 (2002)}{doi:10.1088/1126-6708/2002/12/051}
\arxivlink{hep-th/0210095};
``On integrability of the Yang-Baxter sigma-model,''
\doilink{J. Math. Phys. \textbf{50}, 043508 (2009)}{doi:10.1063/1.3116242}
\arxivlink{0802.3518}.

\bibitem{balog}
J.~Balog, P.~Forgacs, Z.~Horvath and L.~Palla,
``A New family of SU(2) symmetric integrable sigma models,''
\doilink{Phys.\ Lett.\ B {\bf 324}, 403 (1994)}{doi:10.1016/0370-2693(94)90213-5}
\arxivlink{hep-th/9307030}.

\bibitem{DMV}
F.~Delduc, M.~Magro and B.~Vicedo,
``On classical $q$-deformations of integrable sigma-models,''
\doilink{JHEP \textbf{11}, 192 (2013)}{doi:10.1007/JHEP11(2013)192}
\arxivlink{1308.3581};
``Integrable double deformation of the principal chiral model,''
\doilink{Nucl. Phys. B \textbf{891}, 312-321 (2015)}{doi:10.1016/j.nuclphysb.2014.12.018}
\arxivlink{1410.8066}.

\bibitem{today}
F.~Delduc, S.~Lacroix, K.~Sfetsos and K.~Siampos,
``RG flows of integrable $\sigma$-models and the twist function,''
\doilink{JHEP {\bf 2102}, 065 (2021)}{doi:10.1007/JHEP02(2021)065}
\arxivlink{2010.07879}.
  
  
\bibitem{mt} R.~R.~Metsaev and A.~A.~Tseytlin, 
``Order alpha-prime (two loop) equivalence of the string equations of motion and the sigma model Weyl invariance conditions: dependence on the dilaton and the antisymmetric tensor,''
\doilink{Nucl.\ Phys.\ B {\bf 293}, 385 (1987)}{10.1016/0550-3213(87)90077-0};
``Two loop $\beta$-function for the generalized bosonic sigma model,''
\doilink{Phys.\ Lett.\ B {\bf 191}, 354 (1987)}{doi:10.1016/0370-2693(87)90622-8}.

\bibitem{Bos}
M.~Bos,
``Dimensional Regularization in the Wess-Zumino-Witten Model,''
\doilink{Phys. Lett. B \textbf{189}, 435-441 (1987)}{doi:10.1016/0370-2693(87)90656-3}.

\bibitem{Hull} 
C.~M.~Hull and P.~K.~Townsend,
``The Two Loop Beta Function for $\sigma$ Models With Torsion,''
\doilink{Phys.\ Lett.\ B {\bf 191}, 115 (1987)}{doi:10.1016/0370-2693(87)91331-1}.

S.~V.~Ketov,
``Two Loop Calculations in $\sigma$ Model With Torsion,''
\doilink{Nucl.\ Phys.\ B {\bf 294}, 813 (1987)}{doi:10.1016/0550-3213(87)90609-2}.

D.~Zanon,
``Two Loop Beta Functions and Low-energy String Effective Action for the Two-dimensional Bosonic Nonlinear $\sigma$ Model With a {Wess-Zumino}-witten Term,''
\doilink{Phys.\ Lett.\ B {\bf 191}, 363 (1987)}{doi:10.1016/0370-2693(87)90623-X}.

\bibitem{Bap}
M.~Bos,
``An Example of Dimensional Regularization With Antisymmetric Tensors,''
\doilink{Annals Phys. \textbf{181}, 177 (1988)}{doi:10.1016/0003-4916(88)90164-9}.

\bibitem{KZ}
V.~G.~Knizhnik and A.~B.~Zamolodchikov,
``Current Algebra and Wess-Zumino Model in Two-Dimensions,''
\doilink{Nucl. Phys. B \textbf{247}, 83 (1984)}{doi:10.1016/0550-3213(84)90374-2}.

\bibitem{shifman}
D.~Schubring and M.~Shifman,
``Sigma model on a squashed sphere with a Wess-Zumino term,''
\doilink{Phys.\ Rev.\ D {\bf 103}, no. 2, 025016 (2021)}{doi:10.1103/PhysRevD.103.025016}
\arxivlink{2002.04696}.

\bibitem{romans}
D.~N.~Page and C.~N.~Pope,
``Which Compactifications of $D=11$ Supergravity Are Stable?,''
\doilink{Phys.\ Lett.\  {\bf 144B}, 346 (1984)}{doi:10.1016/0370-2693(84)91275-9}.

L.~J.~Romans,
``New Compactifications of Chiral N=2,  d=10 Supergravity,''
\doilink{Phys. Lett. B \textbf{153}, 392 (1985)}{doi:10.1016/0370-2693(85)90479-4}.

P.~Candelas and X.~C.~de la Ossa,
``Comments on Conifolds,''
\doilink{Nucl. Phys. B \textbf{342}, 246 (1990)}{doi:10.1016/0550-3213(90)90577-Z}.

I.~R.~Klebanov and E.~Witten,
``Superconformal field theory on three-branes at a Calabi-Yau singularity,''
\doilink{Nucl. Phys. B \textbf{536}, 199 (1998)}{doi:10.1016/S0550-3213(98)00654-3}
\arxivlink{hep-th/9807080}.

\bibitem{GMM}
E.~Guadagnini, M.~Martellini and M.~Mintchev,
``Scale invariance of sigma models  on homogeneous spaces,''
\doilink{Phys. Lett. B \textbf{194}, 69 (1987)}{doi:10.1016/0370-2693(87)90771-4}.

E.~Guadagnini,
``Current Algebra in $\sigma$ Models on Homogeneous Spaces,''
\doilink{Nucl. Phys. B \textbf{290}, 417 (1987)}{doi:10.1016/0550-3213(87)90195-7}.


\bibitem{PZT}
L.~A.~Pando Zayas and A.~A.~Tseytlin,
``Conformal sigma models for a class of T(p,q) spaces,''
\doilink{Class. Quant. Grav. \textbf{17}, 5125-5131 (2000)}{doi:10.1088/0264-9381/17/24/312}
\arxivlink{hep-th/0007086}.

\bibitem{maillet}
J.~M.~Maillet,
``Kac-Moody Algebra and Extended Yang-Baxter Relations in the O($N$) Nonlinear $\sigma$ Model,''
\doilink{Phys. Lett. B \textbf{162}, 137 (1985)}{doi:10.1016/0370-2693(85)91075-5};
``New Integrable Canonical Structures in Two-dimensional Models,''
\doilink{Nucl. Phys. B \textbf{269}, 54 (1986)}{doi:10.1016/0550-3213(86)90365-2}.

\bibitem{LMV}
S.~Lacroix, M.~Magro and B.~Vicedo,
``Local charges in involution and hierarchies in integrable sigma-models,''
\doilink{JHEP \textbf{09}, 117 (2017)}{doi:10.1007/JHEP09(2017)117}
\arxivlink{1703.01951}.

\bibitem{Sfetsos:2013wia} 
K.~Sfetsos,
``Integrable interpolations: From exact CFTs to non-Abelian T-duals,''
\doilink{Nucl.\ Phys.\ B {\bf 880}, 225 (2014)}{doi:10.1016/j.nuclphysb.2014.01.004}
\arxivlink{1312.4560}.

T.~J.~Hollowood, J.~L.~Miramontes and D.~M.~Schmidtt,
``Integrable Deformations of Strings on Symmetric Spaces,''
\doilink{JHEP \textbf{11}, 009 (2014)}{doi:10.1007/JHEP11(2014)009}
\arxivlink{1407.2840}.

\bibitem{sfetsos} G.~Georgiou and K.~Sfetsos,
``A new class of integrable deformations of CFTs,''
\doilink{JHEP {\bf 1703}, 083 (2017)}{10.1007/JHEP03(2017)083}
\arxivlink{1612.05012};
``Integrable flows between exact CFTs,''
\doilink{JHEP {\bf 1711}, 078 (2017)}{10.1007/JHEP11(2017)078}
\arxivlink{1707.05149}.

G.~Georgiou, E.~Sagkrioti, K.~Sfetsos and K.~Siampos,
``Quantum aspects of doubly deformed CFTs,''
\doilink{Nucl.\ Phys.\ B {\bf 919}, 504 (2017)}{10.1016/j.nuclphysb.2017.04.004}
\arxivlink{1703.00462}.

G.~Georgiou, K.~Sfetsos and K.~Siampos,
``Double and cyclic $\lambda$-deformations and their canonical equivalents,''
\doilink{Phys.\ Lett.\ B {\bf 771}, 576 (2017)}{10.1016/j.physletb.2017.06.007}
\arxivlink{1704.07834}.

\bibitem{s19}
G.~Georgiou, E.~Sagkrioti, K.~Sfetsos and K.~Siampos,
``An exact symmetry in $\lambda$-deformed CFTs,''
\doilink{JHEP \textbf{01}, 083 (2020)}{doi:10.1007/JHEP01(2020)083}
\arxivlink{1911.02027}.

\bibitem{ss}
K.~Sfetsos and K.~Siampos,
``Gauged WZW-type theories and the all-loop anisotropic non-abelian Thirring model,''
\doilink{Nucl.\ Phys.\ B {\bf 885}, 583 (2014)}{10.1016/j.nuclphysb.2014.06.012},
\arxivlink{1405.7803}.

\bibitem{ah}
C.~Appadu and T.~J.~Hollowood,
``Beta function of $k$ deformed $AdS_5 \times S^5$ string theory,''
\doilink{JHEP {\bf 1511}, 095 (2015)}{10.1007/JHEP11(2015)095},
\arxivlink{1507.05420}.

\bibitem{HS}
C.~M.~Hull and B.~J.~Spence,
``The Gauged Nonlinear $\sigma$ Model With {Wess-Zumino} Term,''
\doilink{Phys. Lett. B \textbf{232}, 204-210 (1989)}{doi:10.1016/0370-2693(89)91688-2}.

\bibitem{witten}
E.~Witten,
``On Holomorphic factorization of WZW and coset models,''
\doilink{Commun. Math. Phys. \textbf{144}, 189-212 (1992)}{doi:10.1007/BF02099196}.

\bibitem{belo} 
V.~V.~Belokurov and P.~M.~de Barrush Pasheku Seara de Sa,
``Ultraviolet finiteness of the Wess-Zumino-Witten gauge model on homogeneous manifolds,''
Moscow Univ.\ Phys.\ Bull.\  {\bf 45N3}, 13 (1990)
[Vestn.\ Mosk.\ Univ.\ Fiz.\ Astron.\  {\bf 31N3}, 13 (1990)].

\bibitem{BBS}
K.~Bardakci, L.~M.~Bernardo and N.~Sochen,
``Integrable generalized Thirring model,''
\doilink{Nucl. Phys. B \textbf{487}, 513-525 (1997)}{doi:10.1016/S0550-3213(96)00715-8}
\arxivlink{hep-th/9607018}.

\bibitem{bpz}
P.~Basu and L.~A.~Pando Zayas,
``Chaos rules out integrability of strings on AdS$_5 \times T^{1,1}$,''
\doilink{Phys. Lett. B \textbf{700}, 243-248 (2011)}{doi:10.1016/j.physletb.2011.04.063}
\arxivlink{1103.4107};
``Analytic Non-integrability in String Theory,''
\doilink{Phys. Rev. D \textbf{84}, 046006 (2011)}{doi:10.1103/PhysRevD.84.046006}
\arxivlink{1105.2540}.

\bibitem{yosh}
I.~Kawaguchi, D.~Orlando and K.~Yoshida,
``Yangian symmetry in deformed WZNW models on squashed spheres,''
\doilink{Phys.\ Lett.\ B {\bf 701}, 475 (2011)}{doi:10.1016/j.physletb.2011.06.007}
\arxivlink{1104.0738}.
  
I.~Kawaguchi and K.~Yoshida,
``A deformation of quantum affine algebra in squashed Wess-Zumino-Novikov-Witten models,''
\doilink{J.\ Math.\ Phys.\  {\bf 55}, 062302 (2014)}{doi:10.1063/1.4880341}
\arxivlink{1311.4696}.

S.~Demulder, S.~Driezen, A.~Sevrin and D.~C.~Thompson,
``Classical and Quantum Aspects of Yang-Baxter Wess-Zumino Models,''
\doilink{JHEP {\bf 1803}, 041 (2018)}{doi:10.1007/JHEP03(2018)041}
\arxivlink{1711.00084}.

\bibitem{Tseytlin:1991wr} 
A.~A.~Tseytlin,
``Duality and dilaton,''
\doilink{Mod.\ Phys.\ Lett.\ A {\bf 6}, 1721 (1991)}{10.1142/S021773239100186X}, \\
{\small \url{https://www.academia.edu/40796716/Tseytlin1991_duality_dilaton_MPLA} }

P.~E.~Haagensen and K.~Olsen,
``T duality and two loop renormalization flows,''
\doilink{Nucl.\ Phys.\ B {\bf 504}, 326 (1997)}{doi:10.1016/S0550-3213(97)00496-3}
\arxivlink{hep-th/9704157}.

N.~Kaloper and K.~A.~Meissner,
``Duality beyond the first loop,''
\doilink{Phys. Rev. D \textbf{56}, 7940-7953 (1997)}{doi:10.1103/PhysRevD.56.7940}
\arxivlink{hep-th/9705193}.

\bibitem{bu}
T.~H.~Buscher,
``Path Integral Derivation of Quantum Duality in Nonlinear Sigma Models,''
\doilink{Phys.\ Lett.\ B {\bf 201}, 466 (1988)}{doi:10.1016/0370-2693(88)90602-8}.

A.~S.~Schwarz and A.~A.~Tseytlin,
``Dilaton shift under duality and torsion of elliptic complex,''
\doilink{Nucl.\ Phys.\ B {\bf 399}, 691 (1993)}{doi:10.1016/0550-3213(93)90514-P}
\arxivlink{hep-th/9210015}.
  
\bibitem{hassler}
F.~Hassler and T.~B.~Rochais,
``O($D$,$D$)-covariant two-loop $\beta$-functions and Poisson-Lie T-duality,''
\arxivlink{2011.15130}.

F.~Hassler,
``RG flow of integrable $\mathcal{E}$-models,''
\arxivlink{2012.10451}.

\bibitem{Tseytlin:1992pq} A.~A.~Tseytlin,
``String vacuum backgrounds with covariantly constant null Killing vector and 2-d quantum gravity,''
\doilink{Nucl. Phys. B \textbf{390}, 153-172 (1993)}{10.1016/0550-3213(93)90389-7}
\arxivlink{hep-th/9209023};
``Finite $\s$-models and exact string solutions with Minkowski signature metric,''
\doilink{Phys.\ Rev.\ D {\bf 47}, 3421 (1993)}{10.1103/PhysRevD.47.3421}
\arxivlink{hep-th/9211061}.

\bibitem{HLT3}
B.~Hoare, N.~Levine and A.~A.~Tseytlin,
``Sigma models with local couplings: a new integrability -- RG flow connection,''
\doilink{JHEP \textbf{11}, 020 (2020)}{10.1007/JHEP11(2020)020}
\arxivlink{2008.01112}.

\end{thebibliography}
\end{document}